\documentclass[reprint,superscriptaddress,longbibliography]{revtex4-1}
\usepackage{graphicx}
\usepackage{upgreek}
\usepackage{amsmath}
\usepackage{bbding}
\usepackage{float}
\newcommand{\markup}[1]{#1}

\begin{document}

%\preprint{}

\title{Accurate metasurface synthesis incorporating near-field coupling effects}

\author{A. E. Olk}
\email[]{andreas.olk@iee.lu}
%\thanks{}
\affiliation{School of Engineering and Information Technology, The University of New South Wales, Canberra, Australia}
\affiliation{IEE S.A., ZAE Weiergewan, 11 Rue Edmond Reuter, L-5326 Contern, Luxembourg}

\author{D. A. Powell}
\affiliation{School of Engineering and Information Technology, The University of New South Wales, Canberra, Australia}

\date{\today}

\begin{abstract}
One of the most promising metasurface architectures for the microwave and terahertz frequency ranges consists of three patterned metallic layers separated by dielectrics. 
Such metasurfaces are well suited to planar fabrication techniques and their synthesis is facilitated by modelling them as impedance sheets separated by transmission lines. 
We show that this model can be significantly inaccurate in some cases, due to near-field coupling between metallic layers. This problem is particularly severe for higher frequency designs, where fabrication tolerances prevent the patterns from being highly-subwavelength in size.
Since the near-field coupling is difficult to describe analytically, correcting for it in a design typically requires numerical optimization. %
We propose an extension of the widely used equivalent-circuit model to incorporate near-field coupling and show that the extended model can predict the scattering parameters of a metasurface accurately. %
Based on our extended model, we introduce an improved metasurface synthesis algorithm that gives physical insight to the problem and efficiently compensates for the perturbations induced by near-field coupling. %
Using the proposed algorithm, a Huygens metasurface for beam refraction is synthesized showing a performance close to the theoretical efficiency limit despite the presence of strong near-field coupling.

\end{abstract}

%\keywords{}

\maketitle

\section{Introduction \label{sec:intro}}

Metasurfaces are thin sheets of subwavelength resonators which have emerged as a versatile platform for wavefront manipulation, addressing applications from microwave to visible frequencies \cite{yuFlatOpticsDesigner2014,dingGradientMetasurfacesReview2018,chenHuygensMetasurfacesMicrowaves2018}. Where high transmission efficiency is required, Huygens metasurfaces are widely used as they feature equal electric and magnetic dipole responses to suppress spurious reflections \cite{pfeifferMetamaterialHuygensSurfaces2013a,monticoneFullControlNanoscale2013,deckerHighEfficiencyDielectricHuygens2015}. 
Different metasurface architectures and corresponding synthesis methods have been proposed within the last two decades \cite{dingGradientMetasurfacesReview2018}. 
%All-dielectric metasurfaces have been shown to be efficient for applications at optical frequencies due to their low losses \cite{deckerHighEfficiencyDielectricHuygens2015,shalaevHighEfficiencyAllDielectricMetasurfaces2015,paniagua-dominguezMetalensNearUnityNumerical2018}. 
For microwave to terahertz frequencies, most reported work is based on planar stacks of three patterned metallic layers, separated by dielectric substrates, as this architecture is compatible with planar circuit manufacturing techniques \cite{chenHuygensMetasurfacesMicrowaves2018,epsteinHuygensMetasurfacesEquivalence2016,pfeifferBianisotropicMetasurfacesOptimal2014}. In addition to Huygens metasurfaces, the same approach has been used to realize bianisotropic metasurfaces, which are more efficient for very large refraction angles \cite{wongReflectionlessWideAngleRefracting2016,ColeRefractionefficiencyHuygens2018,fathnanBandwidthSizeLimits2018}. 

In synthesizing metasurfaces, it is often necessary to assume that meta-atoms are small compared to the wavelength. This has motivated the development of deeply subwavelength meta-atoms \cite{wongDesignUnitCells2014,epsteinHuygensMetasurfacesEquivalence2016}, with feature sizes on the order of $\lambda/200$. For metasurfaces operating at W-band frequencies (75 - 110\,GHz) and higher, the resulting feature sizes are too small for conventional printed circuit board fabrication techniques, so simpler geometries such as dogbone \cite{lavigneSusceptibilityDerivationExperimental2018,capolinoEquivalentTransmissionLine2013,rabinovichAnalyticalDesignPrintedCircuitBoard2018} must be used. This frequency range is of interest for applications such as wireless communication \cite{rappaportOverviewMillimeterWave2017} and radar \cite{patoleAutomotiveRadarsReview2017,haschMillimeterWaveTechnologyAutomotive2012}, and metasurfaces should enable devices such as efficient transmitarray antennas \cite{pfeifferMillimeterWaveTransmitarraysWavefront2013,jiangHighlyEfficientBroadband2018}. Despite the high potential for applications, to the best of our knowledge, there is only one work reporting transmissive refracting metasurfaces operating at W-band frequencies \cite{pfeifferMillimeterWaveTransmitarraysWavefront2013}. \markup{Efficient modelling of near-field interaction would enable metasurfaces to be synthesized with simpler geometries, which are feasible to fabricate for this frequency range}.

In modelling multilayered printed circuit metasurfaces, each metallic layer is represented by an equivalent surface impedance and the dielectric layers in between are treated as transmission line sections (See Fig.~\ref{fig:3l_intro}) \cite{epsteinArbitraryPowerConservingField2016,pfeifferMillimeterWaveTransmitarraysWavefront2013,monticoneFullControlNanoscale2013}. 
The advantage of this model is that it yields expressions for the required surface impedance of each metallic layer, greatly simplifying the process of designing the metasurface. The transmission line model accounts for the propagation of the fundamental Floquet harmonic within a periodic metasurface, neglecting all higher-order harmonics, since they are evanescent. Neglecting evanescent waves is accurate for structures with large separation relative to the wavelength, however in metasurfaces it is usually required that the separation between the metallic layers is small. Thus near-field coupling between the layers is neglected. As we show here, this approximation is significantly inaccurate in cases of practical interest, and can lead to performance degradation. Therefore, accurate modelling and synthesis of \markup{transmissive} metasurfaces must take account of near-field effects between metallic layers. We note that the near-field coupling effects that are investigated here are not related to the evanescent waves (or auxiliary fields) deliberately introduced in Ref.~\onlinecite{epsteinSynthesisPassiveLossless2016}.

Early works on near-field interaction in metamaterials aimed to model the propagation of magneto-inductive waves in arrays of split ring resonators \cite{Symstheorymetamaterialsbased2005}, to analyze tunability through modifying lattice parameters \cite{powellMetamaterialTuningManipulation2010} and to analyze optical activity of twisted dimers  \cite{LiuMagneticplasmonhybridization2007,liuOpticalActivityCoupling2012}. Furthermore, several recent works propose analytical design approaches for metagratings based on rigorous descriptions of the mutual interaction of individual particles  \cite{chalabiEfficientAnomalousReflection2017,radiMetagratingsLimitsGraded2017,rabinovichAnalyticalDesignPrintedCircuitBoard2018,popovControllingDiffractionPatterns2018,epsteinPerfectAnomalousRefraction2018}. %For instance, Al\`{u} et al.~proposed reflective metagratings with bianisotropic dielectric inclusions based on a synthesis methodology where polarisabilities of individual scatterers are derived \cite{chalabiEfficientAnomalousReflection2017,radiMetagratingsLimitsGraded2017}. Similarly, Popov et al.~derived conditions for mutual impedances of conductor lines, proposing a circuit board compatible metagrating geometry \cite{popovControllingDiffractionPatterns2018}. 
While these works on \markup{near-field coupling in} metagratings provide full insight into the interaction of individual scatterers and demonstrate highly efficient devices, they have been largely restricted to reflective operation at large angles. More recently, near-field effects within and between cells of a metasurface have been investigated in gap plasmon structures \cite{deshpandeDirectCharacterizationNearField2018,sharacImplementationPlasmonicBand2018}.
%to understand the performance degradation based on conventional synthesis approaches. 
%
%Desphande et al.~conducted experimental investigations of near-field coupling for gap plasmon structures utilizing scanning near-field optical microscopy \cite{deshpandeDirectCharacterizationNearField2018}. Based on the outcome, they gave recommendations for the resonator size and density in the specific case of gap plasmon metasurfaces. Additionally, the coupling of nested structures in gap surface plasmon metasurfaces was investigated using band structure analysis \cite{sharacImplementationPlasmonicBand2018}.
%
For other important metasurface geometries, especially those for operation in transmission, near-field coupling is still largely unexplored and achieving the designed transmission response often requires black-box optimization \cite{lavigneSusceptibilityDerivationExperimental2018,ColeRefractionefficiencyHuygens2018} or ad hoc iterative design methods \cite{pfeifferMillimeterWaveTransmitarraysWavefront2013}.
%A general synthesis method for multi-layer metasurfaces that accounts for near-field effects has not been reported.

In this paper, we demonstrate how the model of cascaded impedance sheets can be significantly inaccurate due to near-field coupling between metallic layers. We propose an extension of this model to incorporate near-field coupling and show that the extended model can predict the scattering parameters of metasurfaces accurately. This proposed coupling model does not require the explicit calculation of electromagnetic interaction integrals as in previous works \cite{powellMetamaterialTuningManipulation2010,liuOpticalActivityCoupling2012}. \markup{The inaccuracies caused by near-field coupling can hamper the synthesis of transmissive multilayer metasurfaces}. Therefore, we introduce an improved synthesis algorithm based on the proposed model,
%which allows near-field effects to be taken into account, 
which provides good physical insight and reduced computational effort compared to black-box optimization. The algorithm is applied to a homogeneous Huygens metasurface with dogbone resonators, then to an inhomogeneous metasurface exhibiting anomalous refraction.

%Its efficiency is close to the theoretical limit derived in \cite{epsteinHuygensMetasurfacesEquivalence2016}. The methods shown in this paper are very generic and be applied for the synthesis of highly efficient metasurfaces for frequencies ranging from microwave to terahertz.

\section{Design of metasurfaces with three metallic layers\label{sec:without_nf}}

The model of cascaded impedance sheets is widely used in the literature for the analysis of transmissive metasurfaces \cite{monticoneFullControlNanoscale2013,epsteinArbitraryPowerConservingField2016,pfeifferMillimeterWaveTransmitarraysWavefront2013}. With this model, the properties of each layer are specified based on the scattering properties of the multi-layer stack, allowing the synthesis to be reduced to readily solvable problems. However, the model does not account for near-field coupling between layers, which can lead to significant inaccuracy. In the following, we briefly recapitulate the model and show that neglecting near-field coupling can lead to significant inaccuracy in the design of a homogeneous Huygens metasurface.

\begin{figure}
	\includegraphics[width=.345\textwidth]{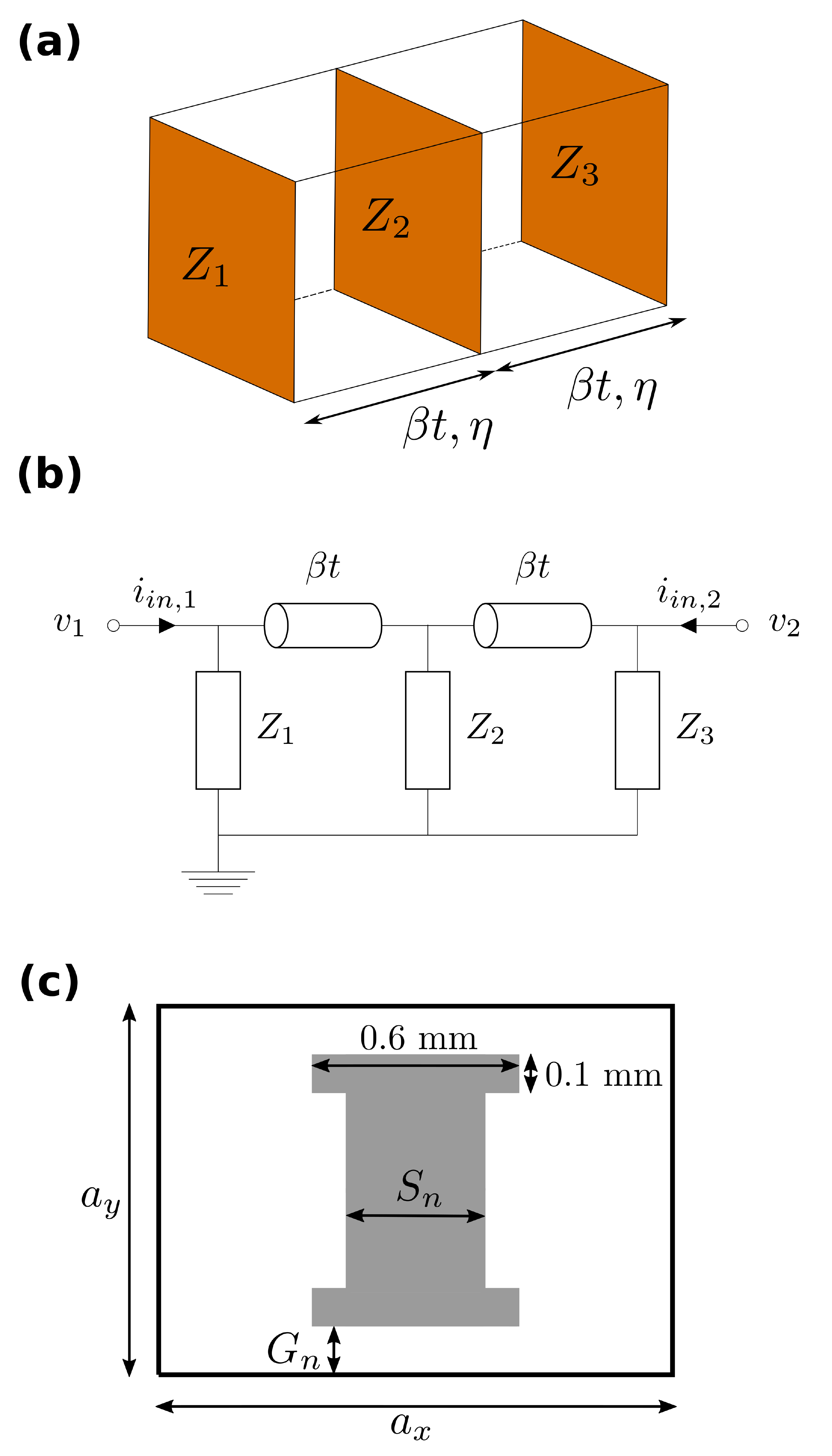}
	\caption{(a) Cascaded impedance sheets representing a multilayer metasurface and (b) corresponding circuit analogy. (c) Metallic pattern of the dogbone metasurface. The parameters used in this section are $G_1=G_3=126\,\upmu$m, $G_2=149\,\upmu$m, $S_1=S_3=300\,\upmu$m and $S_2=400\,\upmu$m. The lateral spacing is $a_x=1.52$\,mm and $a_y=$1.09\,mm.\label{fig:3l_intro}}
\end{figure}

\begin{figure}
	\includegraphics[width=.39\textwidth]{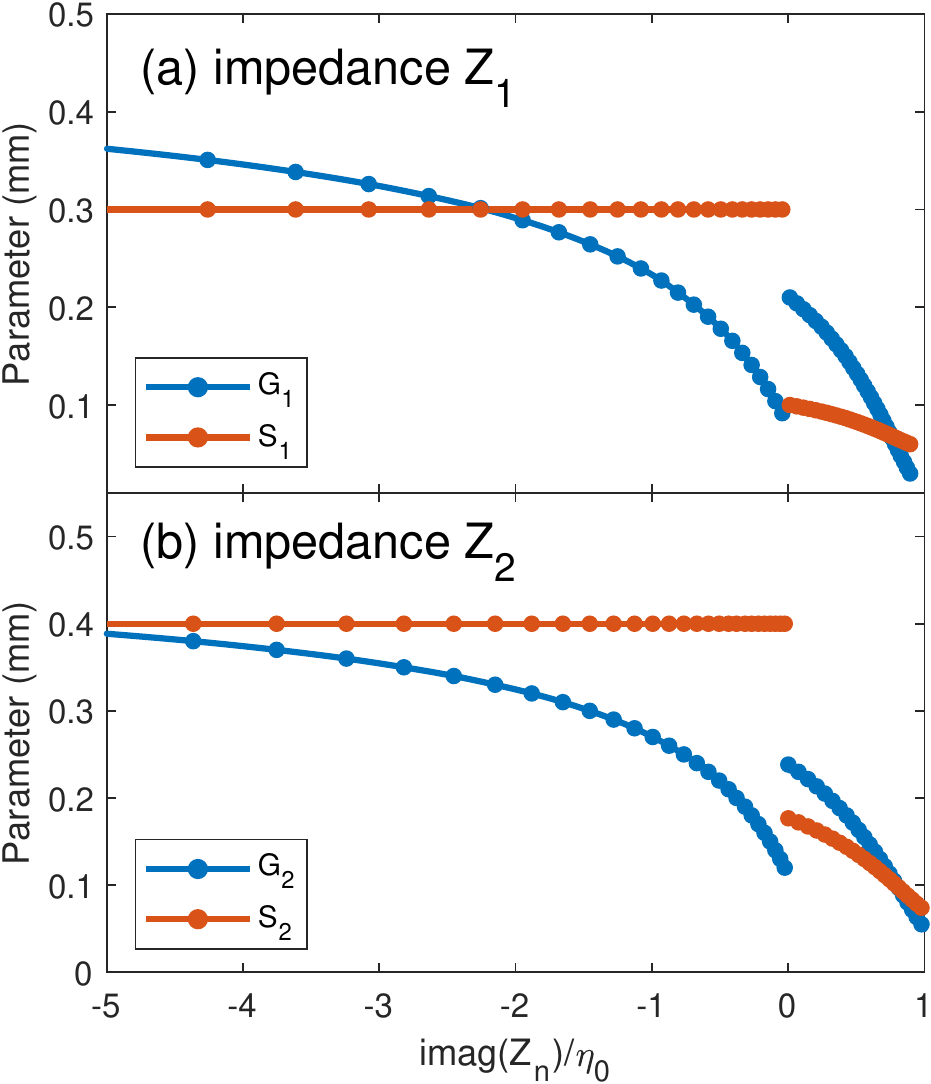}
	\caption{Lookup table for the dogbone cell. The geometrical parameter are related to the shunt impedance of the (a) outer and (b) inner layers. \label{fig:lookup}}
\end{figure}

Each metallic layer is represented by a shunt impedance $Z_n$ and the dielectric layers are represented by a transmission line with propagation constant $\beta$, wave impedance $\eta = \sqrt{\mu / \epsilon}$ and thickness $t$ \cite{tretyakovAnalyticalMethodsApplied2003} as depicted in Figure~\ref{fig:3l_intro}(a). Utilizing a circuit analogy, currents are equivalent to the magnetic field and voltages are equivalent to the electric field \cite{tretyakovAnalyticalMethodsApplied2003}. The transimssion line model accounts for far-field (or first order Floquet) coupling between the metallic-layers. \markup{It is difficult to calculate a meaningful propagation angle within a single meta-atom, therefore within each cell propagation normal to the surface is assumed \cite{epsteinArbitraryPowerConservingField2016}}. The corresponding circuit for the three layer system is shown in Figure~\ref{fig:3l_intro}(b). Using this circuit, the network parameters of each cell can be determined by cascading the individual building blocks in terms of transmission (ABCD) matrices $\mathbf{A_{tot}}=\mathbf{A_{Z1}}\mathbf{A_{tl}}\mathbf{A_{Z2}}\mathbf{A_{tl}}\mathbf{A_{Z3}}$, where the transmission matrix of a shunt impedance $\mathbf{A_{Zn}}$ and a transmission line $\mathbf{A_{tl}}$ are defined in Appendix \ref{sec:matrices}. 
The transmission matrix of the complete cell $\mathbf{A_{tot}}$ is transformed into a scattering matrix $\mathbf{S_{tot}}$. 
In Ref.~\cite{epsteinArbitraryPowerConservingField2016} it was shown how the designed transmission response of the cell leads to specification for each of the three sheet impedances shown in Fig.~\ref{fig:3l_intro}. We perform numerical simulations of the dogbone structure to create a lookup table relating its geometric parameters to the sheet impedance $Z_n$, which is shown in Fig.~\ref{fig:lookup}. We note that the relationship between impedance and geometric parameters differs between the inner and outer layers due to the difference in surrounding dielectric medium.

%Increasing the gap $G_n$ reduces the capacitance and increasing the width $S_n$ of the central connection increases the inductance. 

To apply the model, we synthesize one unit cell of a Huygens metasurface for operation in the millimeter wave band at $f_0=$ 80\,GHz with the metallic layers having the dogbone geometry depicted in Figure~\ref{fig:3l_intro}(c). With the smallest metallic feature size of 0.1\,mm, a substrate thickness of $t=254\,\upmu$m and a permittivity of $\epsilon_r$=3.0, this unit cell design is compatible with standard printed circuit board fabrication while operating at W-band frequencies (similar to Ref.~\onlinecite{pfeifferMillimeterWaveTransmitarraysWavefront2013}). This results in relatively large resonators, with lateral unit cell dimensions of $a \gtrsim \lambda/5$. To allow comparison with theory, we assume lossless dielectric layers, and perfectly conducting metal patterns with zero thickness. Since Huygens metasurfaces are symmetrical, the first and third layer are identical, i.e. $Z_1=Z_3$, $G_1=G_3$ and $S_1=S_3$.

\begin{figure}
	\includegraphics[width=.43\textwidth]{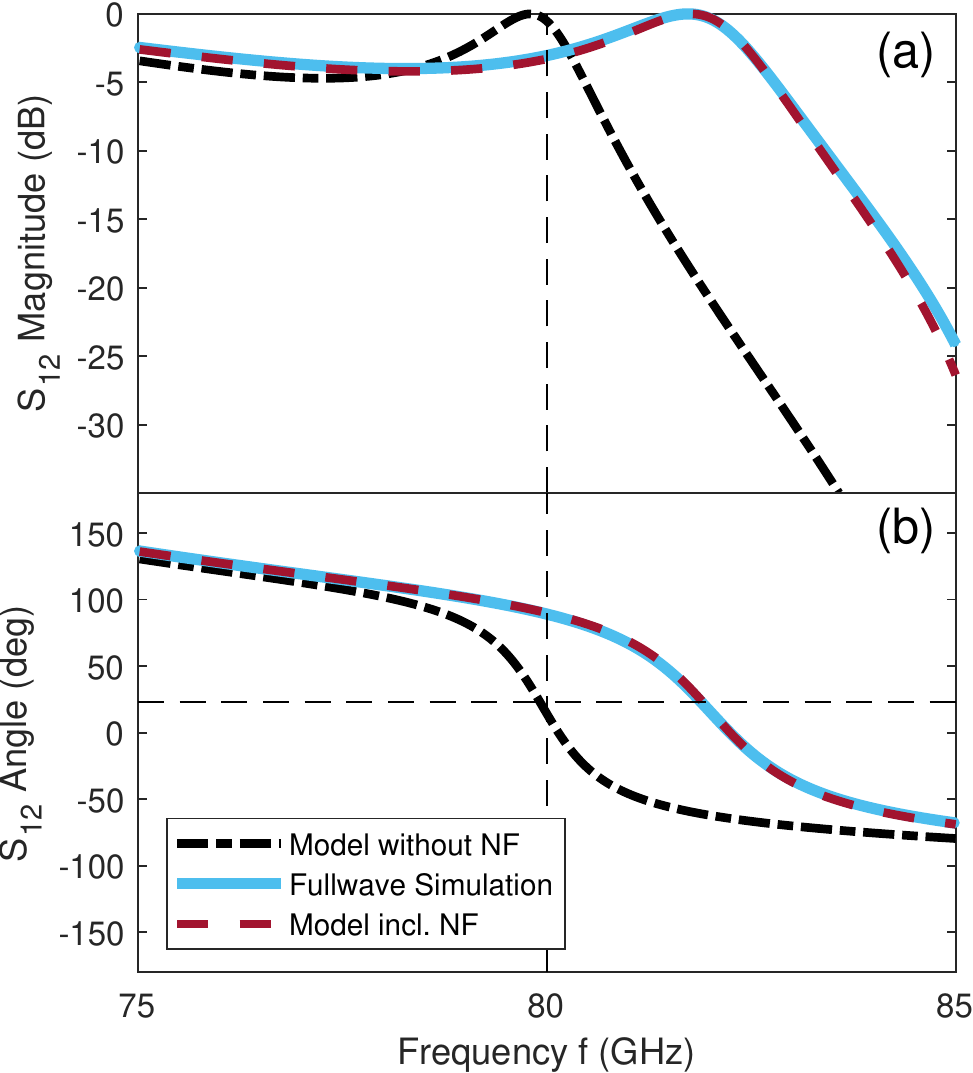}
	\caption{Transmission (a) magnitude and (b) phase of a Huygens metasurface predicted by the existing model (black dashed curve), showing significant disagreement with full-wave simulation (blue curve). Incorporating near-field coupling between layers (red dashed curve) accurately predicts the response. \label{fig:prediction}}
\end{figure}

The metasurface is designed to reach near unity transmission amplitude and a transmission phase response of $\phi_\mathrm{des}=22.9^\circ$ at frequency $f_0$. In Figure~\ref{fig:prediction}, the transmission response of the model (black dashed line) is compared to results from full wave simulation (blue solid line). The model predicts that transmission is close to unity amplitude and meets the expected phase $\phi_\mathrm{des}$ at $f_0$. However, full wave simulation reveals that the transmission maximum and the corresponding phase are significantly shifted to 81.9\,GHz. We attribute this discrepancy to 
near-field coupling (or equivalently higher order Floquet modes), which is to be expected, given that $\beta t \ll 1$. This clearly demonstrates that this widely-used model can fail for some geometries of practical interest, and consequently it is not precise enough for the synthesis of metasurfaces having high efficiency.

%In the following section, we propose an improved model that includes near-field effects between layers, which is capable of predicting this discrepancy accurately (red dashed line). We subsequently introduce a novel optimization algorithm based on the improved model, which corrects these discrepancies efficiently while providing physical insight into the electromagnetic response.

%Within the model, far-field (or first order Floquet) coupling between metallic-layers is represented as waves travelling through the transmission lines. For most metasurfaces however, $\beta t \ll 1$ and near-field (or higher order Floquet) coupling can play a significant role. which are found to exhibit very significant near-field coupling effects. 

%This observation is in agreement with a recent experimental study on gap plasmon metasurfaces, where larger resonators were found to couple stronger to neighboring unit cells \cite{deshpandeDirectCharacterizationNearField2018}. 

\section{Modelling near-field coupling between layers\label{sec:with_nf}}

Here we extend the model of metasurfaces with three metallic layers to incorporate near-field coupling between the layers. We note that the presented model is quite general and could also be applied to bianisotropic metasurfaces, or to designs containing four or more layers \cite{abdo-sanchezLeakyWaveAntennaControlled2018,liuFullyControllablePancharatnamBerry2016,panTrifunctionalMetasurfacesConcept2018,jiangAchromaticElectromagneticMetasurface2018}.

\subsection{Near-field impedance matrix\label{sec:formalism}}

\begin{figure}
    \includegraphics[width=.48\textwidth]{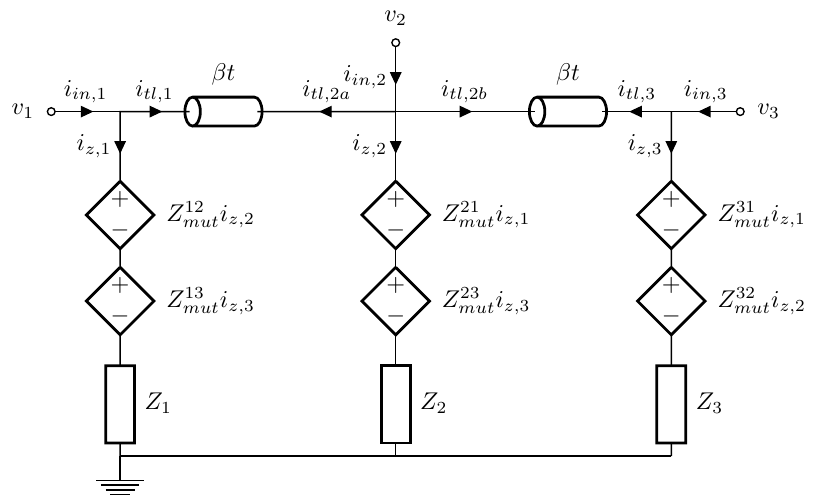}
	\caption{Proposed equivalent circuit model representing a three layer metasurfaces. Inter-layer near-field coupling is included using current controlled voltage sources.\label{fig:circuit}}
\end{figure}

We extend the circuit model shown in Figure~\ref{fig:3l_intro}(b) by accounting for the mutual impedance between each of the metallic layers. As shown in Fig.~\ref{fig:circuit}, this is represented within the equivalent circuit model by current controlled voltage sources. To facilitate our analysis, we include a virtual port at the location of the centre metallic layer. \markup{Consistent with the existing model, we assume that propagation within a single meta-atom is normal to the surface}.

We introduce the near-field impedance matrix $\mathbf{Z_{nf}}$ which gives the near-field relationship between the equivalent voltages $v_n$ and currents $i_{z,n}$ on each of the metallic layers
\begin{equation}
\label{eq:zsheet3}
\begin{bmatrix}
v_{1} \\ v_{2} \\ v_{3}
\end{bmatrix}
=
\begin{bmatrix}
Z_1 & Z_{mut}^{12} & Z_{mut}^{13} \\
Z_{mut}^{21} & Z_2 & Z_{mut}^{23} \\
Z_{mut}^{31} & Z_{mut}^{32} & Z_3 
\end{bmatrix}
\begin{bmatrix}
i_{z,1} \\ i_{z,2} \\ i_{z,3}
\end{bmatrix}.
\end{equation}
The diagonal terms $Z_n$ are the self impedances of layer $n$ and the off diagonal elements are the transfer impedances $Z_{mut}^{lm}$ of the corresponding current controlled voltage sources, hence they represent coefficients that quantify the near-field coupling between layer $l$ and $m$. We consider only reciprocal structures, therefore $\mathbf{Z_{nf}}$ is a symmetric matrix with $Z_{mut}^{lm}=Z_{mut}^{ml}$. Also note that the use of a scalar current $i_{z,n}$ to represent the current flow on a metallic layer is valid so long as only one mode can be excited on it.

To relate $\mathbf{Z_{nf}}$ to the scattering parameters of the entire cell, we first convert it to admittance parameters $\mathbf{Y_{nf}}=\mathbf{Z_{nf}}^{-1}$.
%, $\mathbf{Y_{tot}}$ and $\mathbf{Y_{ff}}$. 
The admittance matrix of the complete cell $\mathbf{Y_{tot}}$ relates the equivalent current $i_{in,n}$ flowing into to each port to the equivalent voltages
\begin{eqnarray}
\begin{bmatrix}
i_{in,1} \\
i_{in,2} \\
i_{in,3} \\
\end{bmatrix}
=
\mathbf{Y_{tot}}
\begin{bmatrix}
v_1 \\
v_2 \\
v_3 \\
\end{bmatrix}.
\end{eqnarray}
Admittance matrix $\mathbf{Y_{ff}}$ represents the far-field contribution of the transmission lines through their equivalent current $i_{tl,n}$.
\begin{eqnarray}
\label{eq:ytl}
\begin{bmatrix}
i_{tl,1} \\
i_{tl,2} \\
i_{tl,3} \\
\end{bmatrix}
=
\mathbf{Y_{ff}}
\begin{bmatrix}
v_1 \\
v_2 \\
v_3 \\
\end{bmatrix},
\end{eqnarray}
where we introduce the total transmission line current at the central node as $i_{tl,2}=i_{tl,2a}+i_{tl,2b}$. 
An analytical expression for $\mathbf{Y_{ff}}$ is given in Appendix~\ref{sec:matrices}. 

Using the relation $i_{in,n}=i_{z,n} + i_{tl,n}$ and Eq.~\eqref{eq:zsheet3} to \eqref{eq:ytl}, we can write 
\begin{equation}
\label{eq:admittanceRel}
\mathbf{Y_{tot}} = \mathbf{Y_{nf}} + \mathbf{Y_{ff}}.
\end{equation}
Using these admittance matrices, near-field and far field contributions to the inter-layer coupling simply sum up and can be linked to the network parameters of the complete cell system $\mathbf{Y_{tot}}$.

As no current can flow through the virtual port 2 of Fig.~\ref{fig:circuit}, we impose the condition $i_{in,2}=0$. This leads to a reduced admittance matrix $\mathbf{Y_{tot}'}$ for the complete cell, defined as
\begin{equation}
\begin{bmatrix}
i_{in,1} \\
i_{in,3} \\
\end{bmatrix}
=
\mathbf{Y_{tot}'}
\begin{bmatrix}
v_1 \\
v_3 \\
\end{bmatrix},
\end{equation}
with its relationship to $\mathbf{Y_{tot}}$ shown in Appendix~\ref{sec:matrices}. 

Given a near-field impedance matrix $\mathbf{Z_{nf}}$, the reduced two port admittance matrix $\mathbf{Y_{tot}'}$ can be determined using equation Eqs.~\eqref{eq:admittanceRel} and \eqref{eq:ytot3to2}. This is then converted to a scattering matrix $\mathbf{S_{tot}}$, to obtain transmission and reflection properties of the metasurface. Note that if the near-field coupling coefficients $Z_{mut}^{lm}$ are set to zero, this formalism is equivalent to the existing model outlined in Section~\ref{sec:without_nf}.

\subsection{Determination of the near-field impedance terms\label{sec:determination}}

To apply the formalism developed in Subsection~\ref{sec:formalism} to metasurface analysis, it is necessary to find the values of elements of $\mathbf{Z_{nf}}$. 
%We have introduced a set of equations that link the near-field impedance matrix $\mathbf{Z_{nf}}$ to the scattering parameters of the metasurface $\mathbf{S_{tot}}$.  
%With these equations, we can determine the scattering matrix of a metasurface $\mathbf{S_{tot}}$ from known $\mathbf{Z_{nf}}$. 
Since $\mathbf{Z_{nf}}$ is a 3x3 matrix and $\mathbf{S_{tot}}$ is a 2x2 matrix, a single far-field scattering simulation yields insufficient information to resolve $\mathbf{Z_{nf}}$. Therefore, we propose a technique to determine the elements of $\mathbf{Z_{nf}}$ successively from full wave simulations of reduced systems containing only two layers. A two layer formalism analogous to Section~\ref{sec:formalism} is applied to these full-wave scattering parameters, which allows the unambiguous extraction of individual elements of $\mathbf{Z_{nf}}$. This two layer formalism is outlined in Appendix~\ref{sec:matrices}.

\begin{figure}[htbp]
	\centering
	\includegraphics[width=.34\textwidth]{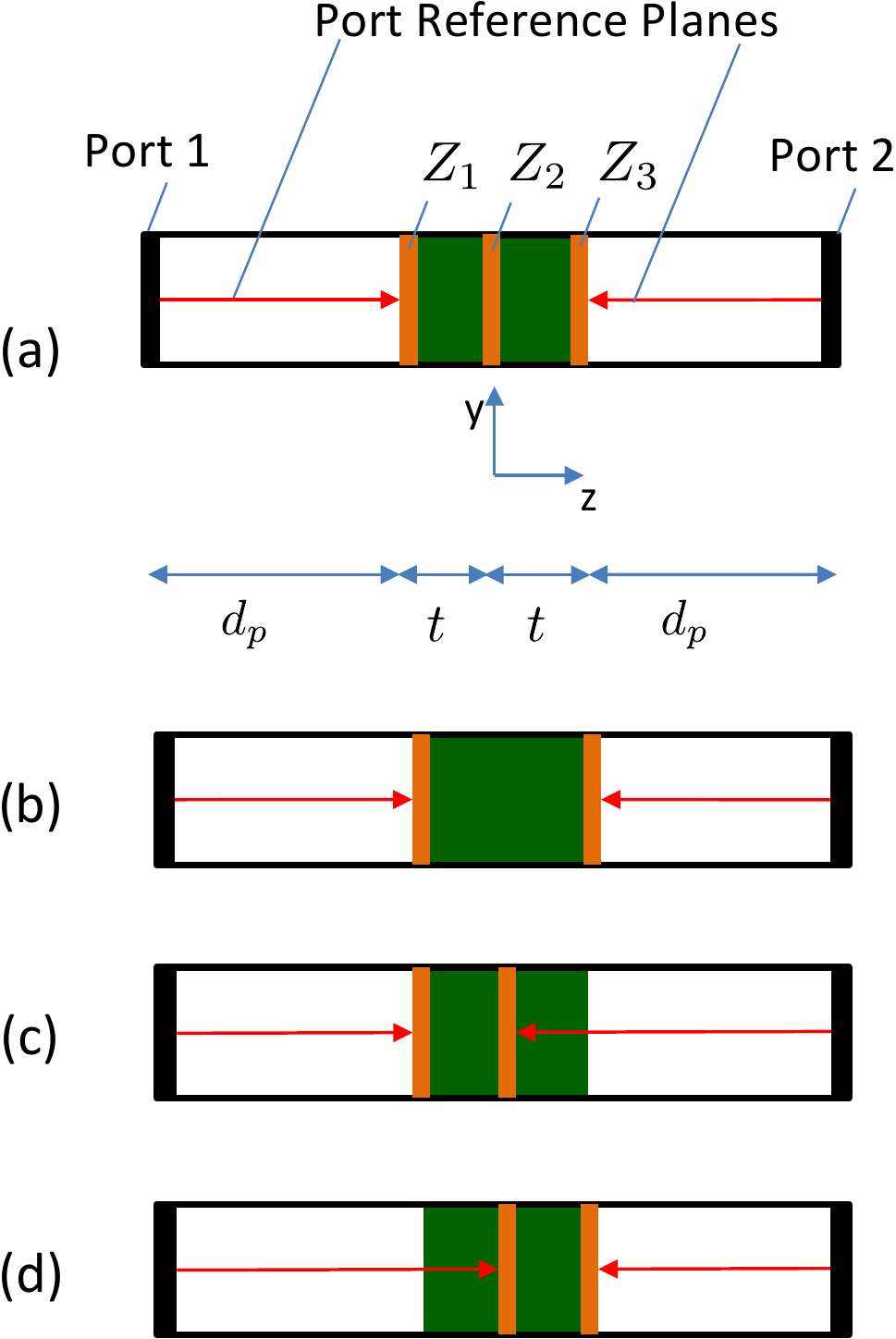}
	\caption{(a) Cross sectional layout of a typical full wave simulation to determine the scattering parameter of a unit cell. (b) - (d) Individual simulations that are required to determine the near-field impedance matrix $\mathbf{Z_{nf}}$.\label{fig:single_sims}}
\end{figure}

A typical configuration for a full wave simulation of the unit cell of a metasurface with three layers is shown in Fig.~\ref{fig:single_sims} (a). The unit cell is centered at $z=0$, periodic boundary conditions apply to all sides and Floquet ports are located at a large distance $d_p$ to the top and bottom of the structure, where the influence of high order Floquet modes is negligible. The reference plane of these ports however needs to be moved back to the top and bottom to $z=\pm t$, respectively, to correct the reflection and transmission phase \cite{epsteinHuygensMetasurfacesEquivalence2016}.

By removing the central metallic layer, as shown in Fig.~\ref{fig:single_sims}(b), the two layer formalism can be applied to determine the self impedance of layers one and three, $Z_1$ and $Z_3$, and their mutual impedance $Z_{mut}^{13}$. To determine $Z_1$, $Z_2$ and $Z_{mut}^{12}$, the configuration from Fig.~\ref{fig:single_sims}(c) is utilized. Here, however, the reference plane of Port 2 needs to be moved to the central metal layer, which can be done by inverting the ABCD matrix of the dielectric layer \cite{pozarMicrowaveEngineering4th2011}. Similarly, $Z_2$, $Z_3$ and $Z_{mut}^{23}$ can be obtained from the configuration from Fig.~\ref{fig:single_sims}~(d). Using this successive approach, the full coupling matrix $\mathbf{Z_{nf}}$ is determined and the three layer system is fully characterized. 

\begin{figure}[htbp]
	\includegraphics[width=.44\textwidth]{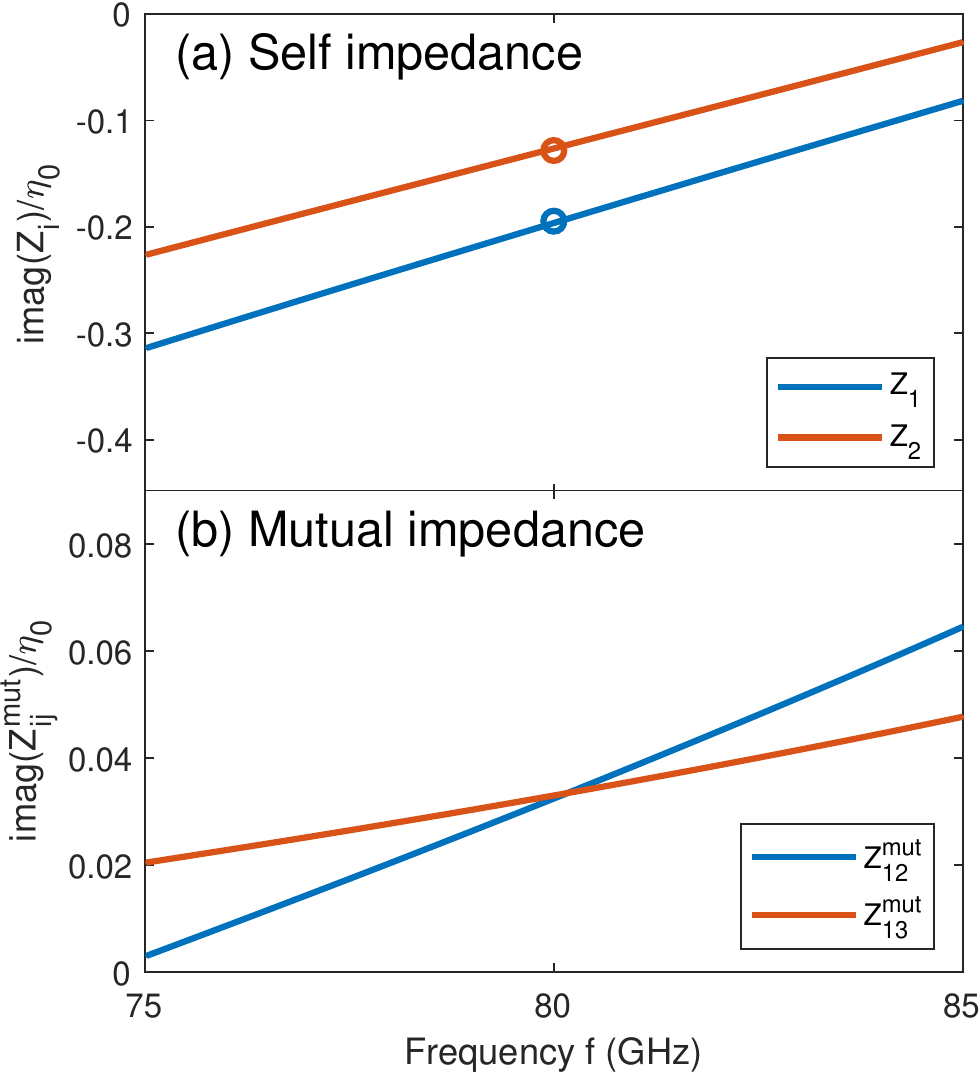}
	\caption{Extracted values of (a) self impedance  imag$(Z_n)$ and (b) mutual impedance  imag$(Z_{mut}^{lm})$ for a three layer dogbone structure.\label{fig:impedances}}	
\end{figure}

We apply this technique to the metasurface considered in Section~\ref{sec:without_nf} to determine the near-field impedance matrix $\mathbf{Z_{nf}}$.
As we consider a Huygens metasurface with $Z_1=Z_3$ and $Z_{mut}^{lm}=Z_{mut}^{ml}$, only two simulations are required to determine the near-field impedance matrix $\mathbf{Z_{nf}}$, since the configurations in Figure~\ref{fig:single_sims}(c) and (d) are equivalent. 
The resulting self impedances $Z_n$ and mutual impedances $Z_{mut}^{lm}$ are shown in Figure~\ref{fig:impedances}. As we consider lossless structures, these functions are purely imaginary to within numerical accuracy ($|\mathrm{Re}(Z)|/\eta_0<10^{-5}$).
%The corresponding real parts in this case are as small as $ |\mathrm{real}(Z)|/\eta_0<10^{-5}$. 
At the design frequency, the mutual impedance $Z_{\mathrm{mut}}^{12}$ and $Z_{mut}^{13}$ are clearly smaller than the self impedance $Z_1$ and $Z_2$, however they are not negligible. The mutual and self impedance terms have the form of a series LC circuit. Using the calculated near-field impedance matrix, the metasurface transmission is plotted in Figure~\ref{fig:prediction} (red dashed line), showing good agreement with the full wave simulation. This result confirms the accuracy of our near-field coupling model, demonstrating why the conventional metasurface synthesis procedure fails in this case.

\section{Metasurface synthesis incorporating near-field coupling}

\subsection{Refinement algorithm for metasurface synthesis\label{sec:algo}}

Having characterized how near-field coupling can render existing metasurface design techniques %\cite{epsteinArbitraryPowerConservingField2016,monticoneFullControlNanoscale2013,pfeifferMillimeterWaveTransmitarraysWavefront2013} 
inaccurate, we present an approach to include it within the metasurface synthesis. Since the near-field coupling cannot be evaluated until a geometry of the metal layer has been specified, this necessarily involves the use of an iterative algorithm. The goal of the algorithm is to find the geometrical parameters of the unit cell that match the target scattering parameters. 
%The problem is formulated in terms of scattering parameters,  however, it can be formulated using any other network parameters.  
In the case of Huygens metasurfaces, it is sufficient to minimize the difference between the realized transmission $S_{21}^\mathrm{re}$ and designed transmission $S_{21}^\mathrm{des}$
\begin{equation}
\text{min}  ( |S_{21}^\mathrm{re}-S_{21}^\mathrm{des}|^2 ).
\end{equation}
The realized transmission $S_{21}^\mathrm{re}$ is calculated from our near-field coupling model with the matrix $\mathbf{Z_{nf}}$. A fundamental assumption of the algorithm is that for small changes of self impedances $Z_n$, the geometric change is small, and the mutual coupling does not change strongly. This allows us to correct only the diagonal elements of the coupling matrix with a factor $\gamma$ at each iteration
\begin{equation}
\label{eq:zsheet3cor}
\mathbf{Z_{nf,p}} = \mathbf{Z_{nf,p-1}}+\gamma \, \text{diag}(Z_{1,p-1},Z_{2,p-1},Z_{3,p-1})
\end{equation} 
where the iteration index is denoted with $p$. The factor $\gamma$ is chosen by minimizing the term $ | S_{12}^\mathrm{re}(\mathbf{Z_{nf,p}}) - S_{12}^\mathrm{des}|^2$ numerically in each iteration. 

Based on the corrected self impedances $Z_{n,p}$, the geometry of the unit cell is updated with new geometrical parameters $G_n$ and $S_n$ using the  lookup table in Figure~\ref{fig:lookup}. If a further iteration is necessary, the mutual impedances are recalculated with full wave simulations of the new geometry. Note that Eq.~\eqref{eq:zsheet3cor} could be modified to adjust outer layer impedances separately from the inner layer ones, however this was found empirically to offer no improvement in convergence.
%We note here, that it is not obvious that changing all self impedances $Z_n$ by the same amount leads to an efficient correction. It was found empirically, that the determination of one single correction factor $\gamma$ as in  is sufficient to compensate for near-field coupling in Huygens metasurfaces.
The optimization process is outlined in full in Algorithm~\ref{algo}.

\newcommand{\tab}{\hspace{.45cm}} 
\newcommand{\vs}{\\[.15cm]}

\newfloat{algo}{t}{lop}
\floatname{algo}{Algorithm }

\begin{algo}
    \caption{Refinement algorithm for metasurface synthesis. \label{algo}}
	\begin{tabular}{p{.46\textwidth}}
	\textbf{Initialize:} $p=0$ \\
	\textbf{Start geometry:} \\
	\tab Calculate self impedances $Z_{n,p=0}$ from model \\
	\tab without near-field coupling and set geometry \\ 
	\tab accordingly \vs
	\textbf{Calculate near-field coupling terms:}\\
	\tab Determine $Z_{mut,p=0}^{12}$,$Z_{mut,p=0}^{13}$ and $Z_{mut,p=0}^{23}$ \\
	\tab with full wave simulations obtaining $\mathbf{Z_{nf,p=0}}$  \vs
	\textbf{Start iterative steps:} $p=1$ \vs
	\textbf{while}($p<p_{max}$) \\
	\tab Initialize coupling matrix according to Eq.~\eqref{eq:zsheet3cor}  \vs
	\tab Calculate correction factor $\gamma$ by minimizing\\
	\tab $ | S_{12}^\mathrm{re}(\mathbf{Z_{nf,p}}) - S_{12}^\mathrm{des}|^2$ to get new $\mathbf{Z_{nf,p}}$ \vs
	\tab Update geometry accordingly \vs
	\tab \textbf{if}$( |\gamma|<\gamma_c)$  	 \textbf{then}\\
	\tab\tab terminate\\
	\tab \textbf{end} \vs
	\tab Recalculate actual $Z_{mut,p}^{12}$,$Z_{mut,p}^{13}$ and $Z_{mut,p}^{23}$ \\
	\tab with current geometry and update $\mathbf{Z_{nf,p}}$ \vs
	\tab p=p+1 \\
	\textbf{end}\\
		
	\end{tabular}
\end{algo}

\begin{figure}[htbp]
	\includegraphics[width=.43\textwidth]{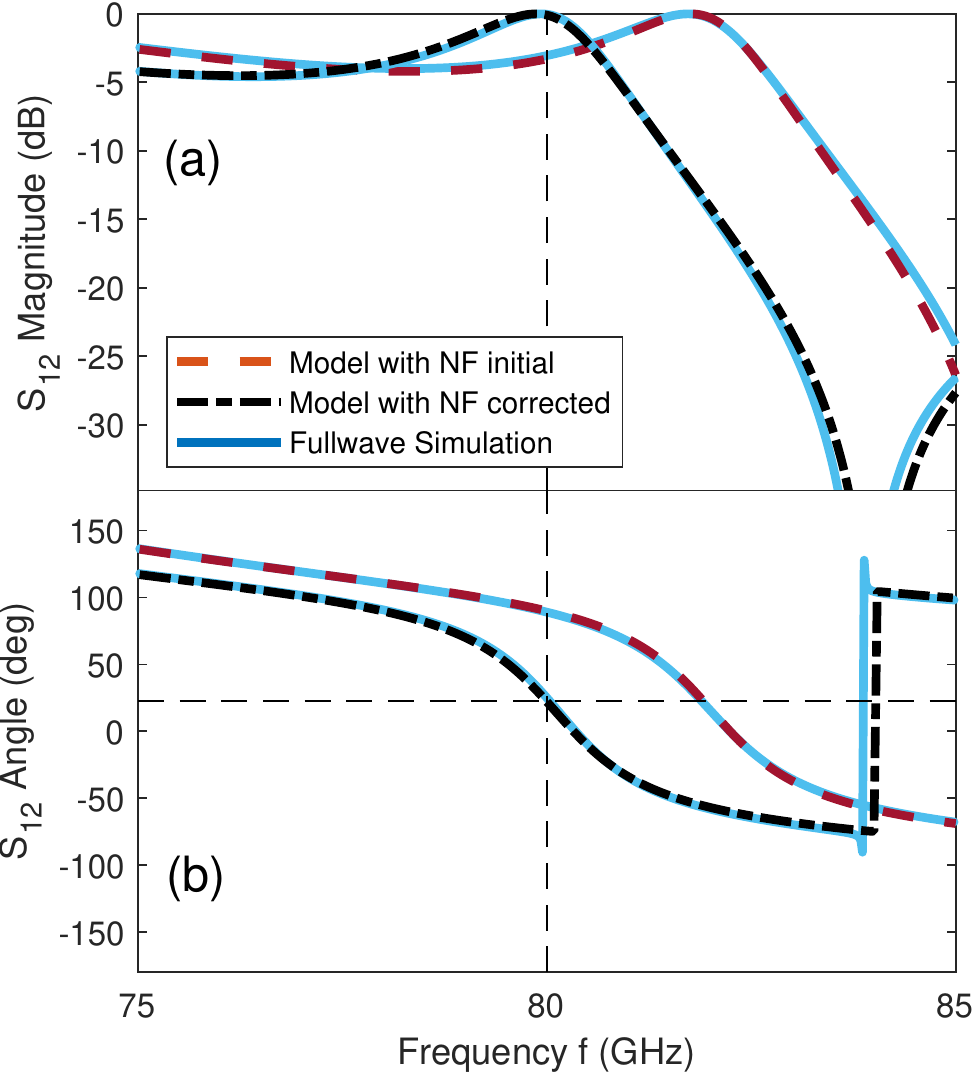}
	\caption{Transmission of Huygens metasurface before and after applying the optimization procedure. (a) Transmission magnitude and (b) phase. The thin dashed lines indicate the designed operating frequency and phase response. \label{fig:supercell_corrected_atom2}}
\end{figure}

The algorithm terminates when the relative change of the self impedances from one iteration to the next is smaller than $\gamma_c$. This convergence parameter was chosen as $\gamma_c=$0.085 for the examples shown in this work. The absolute value of the resulting correction factor for the self impedances $|\gamma|$ ranged between 0.038 and 0.25. The ability of this algorithm to correct the transmission response of a cell is shown in Fig.~\ref{fig:supercell_corrected_atom2}. The red dashed line shows the initial response according to the design procedure outlined in Section~\ref{sec:without_nf}. The black dash-dot curve shows the result of the model after the optimization, and the blue lines show the corresponding results of full-wave simulation. It can be seen that the strong perturbation of the operating frequency has been corrected, and the metasurface now achieves complete transmission and the designed phase value at the operating frequency.

\subsection{Application to metasurface demonstrating anomalous refraction \label{sec:beam_refraction}}

We demonstrate the further utility of our iterative algorithm by applying it to an inhomogeneous Huygens metasurface demonstrating anomalous refraction of a plane wave. The incident angle was chosen $\theta_\mathrm{in}=\,0$ and the refraction angle $\theta_\mathrm{out}=\,55^{\circ}$, leading to a supercell width $d=4.56\,$mm. Each cell is based on the dogbone geometry shown in Fig.~\ref{fig:3l_intro}(a). To minimize the influence of coupling to non-identical neighboring cells which is not described in our model, the lateral spacing $a_x$ was chosen as large as possible. With the supercell width used here, there are three propagating Floquet modes in transmission, $T_0$ and $T_{\pm1}$, and three in reflection $R_0$ and $R_{\pm1}$. To control these 6 degrees of freedom, a minimum of three cells per supercell is required, each having an engineered electric and magnetic response. 

\begin{figure}[htbp]
	\includegraphics[width=.40\textwidth]{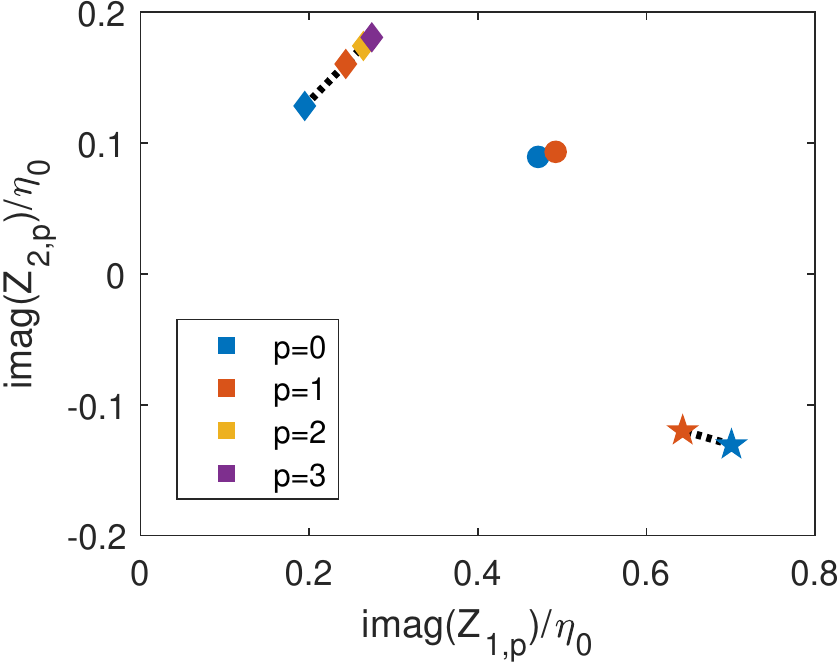}
	\caption{Self impedances of each layer $Z_{n,p}$ at iteration step $p$. The marker shapes $\diamond$, {\large $\circ$} and {\scriptsize \FiveStarOpen} indicate cell 1, 2 and 3 respectively. \label{fig:supercell_synthessis}}
\end{figure}

\markup{The initial sheet impedances $ Z_{n,p=0} $ and the required transmission response for each cell $S_{21}^\mathrm{des}$ were specified according to the design procedure in \cite{epsteinHuygensMetasurfacesEquivalence2016}. With these target values, impedance matching to the refracted wave is stipulated.} The geometry of each cell was synthesized using the algorithm from Section~\ref{sec:algo}. The self impedance $Z_{n,p}$ of each cell $n$ at iteration $p$ is shown in Figure~\ref{fig:supercell_synthessis}. In the case of the first cell, having identical geometry to that was analyzed in Section~\ref{sec:without_nf}, three iterations were necessary to obtain the required scattering properties. For the other two cells, one iteration was enough, due to the lesser influence of near-field coupling in these cases. The resulting geometrical parameters of the supercell are given in Table \ref{tab:supercell_geometry}.

\begin{table}[htbp]
	\begin{center}
	\scalebox{0.9}{
	\begin{tabular}{|c|cccc|cccc|cccc|}
		\hline
		\multicolumn{13}{|c|}{\textbf{Geometry Parameters {[$\mathbf{\upmu}$m]}}}                                \\ \hline
		\multicolumn{1}{|c|}{cell} & \multicolumn{4}{c|}{1} & \multicolumn{4}{c|}{2} & \multicolumn{4}{c|}{3} \\ \hline
		parameter & $S_1$ & $G_1$ & $S_2$ & $G_2$ & $S_1$ & $G_1$ & $S_2$ & $G_2$ & $S_1$ & $G_1$ & $S_2$ & $G_2$ \\ \hline
		before    & 300 & 126 & 400 & 149 & 300 & 204 & 172 & 220 & 300 & 175 & 400 & 138 \\
		after     & 300 & 114 & 400 & 139 & 300 & 210 & 172 & 218 & 300 & 172 & 400 & 137\\ \hline
	\end{tabular}}
	\end{center}
	\caption{Geometric parameters of the cells in the refracting metasurface. \label{tab:supercell_geometry}}
\end{table}

\markup{We note that an alternative approach to correct for near-field coupling is to use black-box optimization methods, which are built in to many commercial simulation packages. Compared to these methods, our proposed algorithm has several advantages. Unlike these black-box optimization methods, our approach gives clear physical insight into the near-field coupling which causes the metasurface design to be non-optimal. In particular, with the quantities $Z_{nm}^{mut}$, near-field coupling effects can be quantified and monitored from iteration to iteration. Additionally, our algorithm is much more computationally efficient, as is demonstrated by the results in Appendix~\ref{sec:app_convergence}, which demonstrates a speed up factor of 3.6 for the design considered. For metasurface designs with a large number of cells, this speed up would also be a significant benefit. Furthermore, using the proposed algorithm, the influence of near-field coupling effects can be corrected without any prior knowledge. Applying black-box optimization methods requires guessing the required geometric parameter changes and setting suitable parameter boundaries.}

\markup{While we considered only a single polarization in this work, we expect that our method could be adapted to tensorial impedances \cite{SelvanayagamDesignMeasurementTensor2016} used to model polarization rotating metasurfaces. We note that for chiral metasurfaces based on coupled high-symmetry layers \cite{Hannam2014}, polarization rotation only occurs when near-field coupling is strong, thus near-field coupling cannot be neglected in equivalent circuit modelling.}

\begin{figure}[htbp]
	\centering
	\includegraphics[width=.44\textwidth]{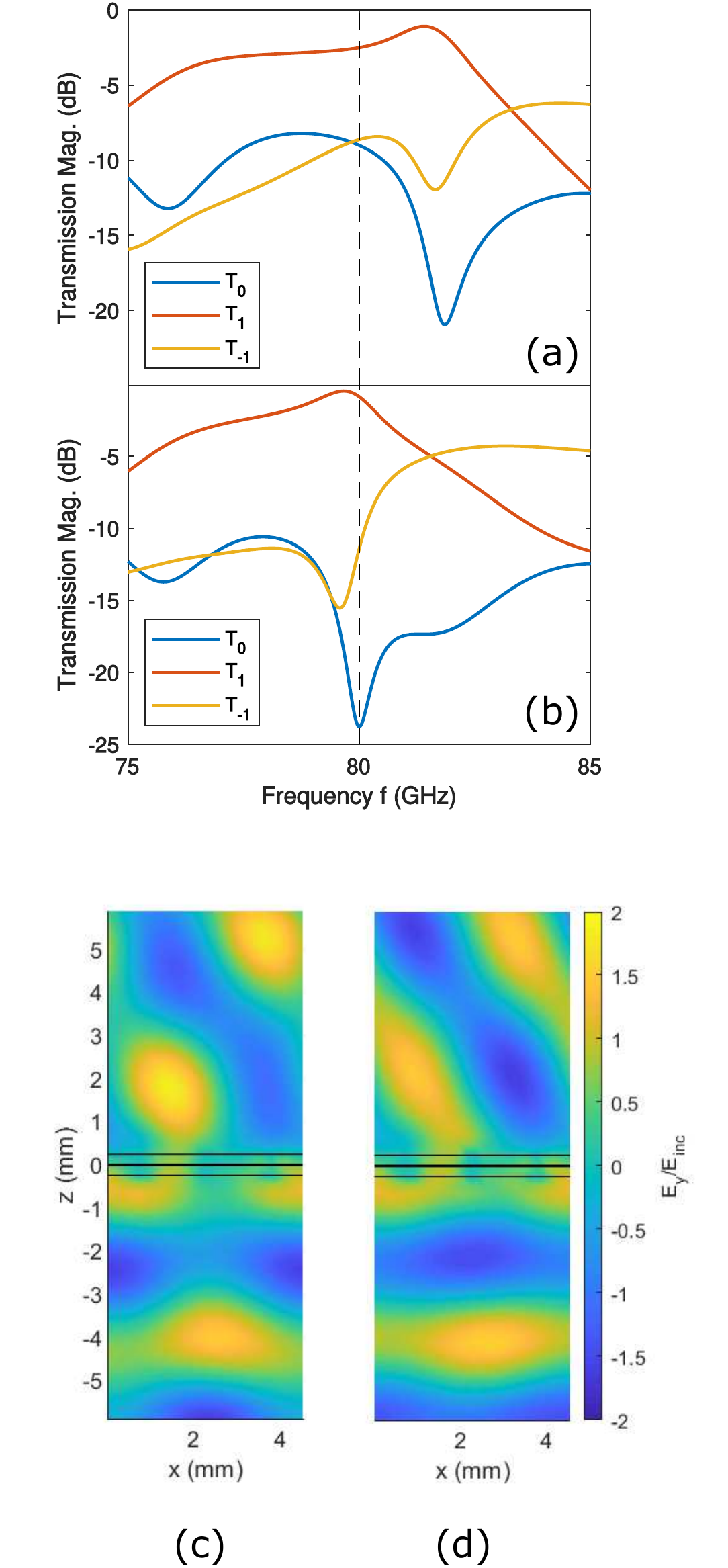}
	\caption{Performance of the beam refraction metasurface. Transmission Floquet modes (a) before and (b) after the model based optimization. Electric field at design frequency $f_0=80\,$GHz (c) before and (d) after the model based optimization.\label{fig:supercell_analysis}}
\end{figure}

The refraction performance of the synthesized metasurface is shown in Figure~\ref{fig:supercell_analysis}, where $T_1$, the transmission coefficient into the designed Floquet mode, should be maximized, and other terms minimized. The performance of the initial geometry (a) is compared with the performance of the geometry obtained by our iterative optimization process (b). For the initial metasurface, the energy refracted into the desired Floquet mode is as low as 56\,\% at the design frequency $f_0=80\,$GHz. The maximum refraction efficiency is 78\,\%, but it is significantly shifted to 81.4\,GHz. The optimized metasurface on the other hand refracts 82\,\% into the desired Floquet mode $T_1$ at the design frequency. In this case, the maximum refraction efficiency is 90\,\% at 79.7\,GHz. We note that the maximum efficiency of the optimized metasurface is close to the theoretical efficiency limit \cite{epsteinHuygensMetasurfacesEquivalence2016} for Huygens metasurfaces, which is 92\,\% for a refraction angle of $55\,^{\circ}$.

\section{Conclusion}

We showed that metasurfaces based on patterned metallic layers separated by dielectric layers can exhibit strong inter-layer near-field coupling. This can lead to significant inaccuracy of existing models used for metasurface synthesis, resulting in poor performance of the metasurfaces. We developed an improved model which accounts for near-field coupling and showed that it closely matches the results of full-wave simulation. This is of particular significance for higher frequency designs, where fabrication tolerances mean that feature sizes are not highly sub-wavelength, leading to strong near-field coupling.

To mitigate the performance degradation induced by near-field coupling, we introduced an iterative algorithm for accurate metasurface synthesis which takes it into account.  The presented algorithm gives more physical insight into the problem than black-box optimization, and reduces computation time significantly. Additionally, it is very generic and can be applied to multilayer metasurfaces in various frequency ranges. As this work introduces an efficient synthesis method for such geometries, it helps to relax fabrication requirements and pave avenues for new metasurface geometries.

\begin{acknowledgments}
	This work was financially supported by the Australian Research Council (Linkage Project LP160100253) and the  Luxembourg Ministry of the Economy (grant CVN 18/17/RED).
\end{acknowledgments}

\appendix

%\section{\label{sec:app_ytl}}

\section{Two and three-port matrix representations}
\label{sec:matrices}

The ABCD matrices of a shunt impedance $\mathbf{A_{Zn}}$ and transmission line $\mathbf{A_{tl}}$ \cite{pozarMicrowaveEngineering4th2011} are given by 

\begin{equation}
\mathbf{A_{Zn}} =
\begin{bmatrix}
1 & 0 \\
1/Z_n & 1 
\end{bmatrix}
\end{equation}
and
\begin{equation}
\mathbf{A_{tl}} =
\begin{bmatrix}
\cos(\beta t) & j \eta \sin(\beta t )\\
j 1/\eta \sin(\beta t ) & \cos(\beta t) 
\end{bmatrix}.
\end{equation}

The admittance matrix $\mathbf{Y_{ff}}$ for the three layer system can be determined analyzing the circuit from Figure~\ref{fig:circuit}~(b) without shunt elements and dependent sources utilizing the definition of an admittance matrix element ${Y_{ff,lm}=\frac{i_{tl,l}}{v_m} \left . \right\vert_{v_{k\neq m}=0}}$ and states 
\begin{equation}
\label{eq:ytl3}
\mathbf{Y_{ff}}
=
\begin{bmatrix}
\frac{-i}{\eta \tan \beta t} & \frac{i}{\eta \sin \beta t} & 0 \\
\frac{i}{\eta \sin \beta t}  & \frac{-2i}{\eta \tan \beta t} & \frac{i}{\eta \sin \beta t} \\
0                                  & \frac{i}{\eta \sin \beta t} & \frac{-i}{\eta \tan \beta t} \\
\end{bmatrix}.
\end{equation}

%\section{\label{sec:app_reducedYtot}}

To reduce the three port equivalent circuit shown in Fig.~\ref{fig:circuit} to a two-port network compatible with circuit simulations, we impose the condition $i_{in,2}=0$. This yields the following expression for the two port matrix $\mathbf{Y_{tot}'}$ in terms of elements of the three port matrix $\mathbf{Y_{tot}}$
\begin{equation}
\label{eq:ytot3to2}
{\small
\mathbf{Y_{tot}'}=
\begin{pmatrix}
Y_{\mathrm{tot},11}-\frac{Y_{\mathrm{tot},21}}{Y_{\mathrm{tot},22}}~Y_{\mathrm{tot},12}  \,\,\,\,& Y_{\mathrm{tot},13}-\frac{Y_{\mathrm{tot},23}}{Y_{\mathrm{tot},22}}~Y_{\mathrm{tot},12} \\[1em]
Y_{\mathrm{tot},31}-\frac{Y_{\mathrm{tot},21}}{Y_{\mathrm{tot},22}}~Y_{\mathrm{tot},32}  \,\,\,\,& Y_{\mathrm{tot},33}-\frac{Y_{\mathrm{tot},23}}{Y_{\mathrm{tot},22}}~Y_{\mathrm{tot},32}
\end{pmatrix}.
}
\end{equation}

%\section{\label{sec:app_2lform}}

\begin{figure}[htbp]
    \includegraphics[width=0.28\textwidth]{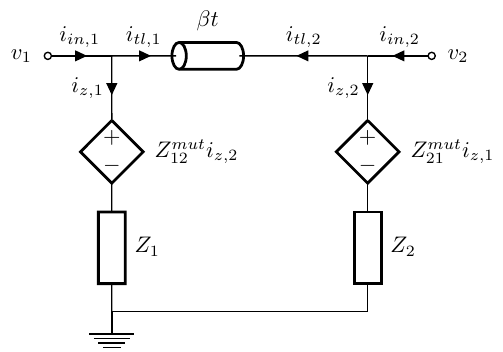}
	\caption{Equivalent circuit for a two layer structure. \label{fig:2l_circuit}}
\end{figure}

Equations \eqref{eq:zsheet3} to \eqref{eq:admittanceRel} can be defined analogously for the two layer system using the 2x2 matrices $\mathbf{Y_{tot}^2}$,$\mathbf{Y_{nf}^2}$ and $\mathbf{Y_{ff}^2}$. In the corresponding circuit model in Figure~\ref{fig:2l_circuit}, the admittance matrix for the transmission line is
\begin{equation}
\mathbf{Y_{ff}^2}=
\begin{pmatrix}
\frac{-j}{\eta \tan \beta t} & \frac{j}{\eta \sin \beta t}  \\[1em]
\frac{j}{\eta \sin \beta t} & \frac{-j}{\eta \tan \beta t}  \\[1em]
\end{pmatrix}.
\end{equation}
With known admittance matrix $\mathbf{Y_{tot}^2}$, we can determine $\mathbf{Z_{nf}^2}=\mathbf{Y_{nf}^2}^{-1}$ without any ambiguity. The coefficients of $\mathbf{Z_{nf}^2}$ for configuration (b) to (d) in Figure~\ref{fig:single_sims} are then used subsequently to build the 3x3 matrix $\mathbf{Z_{nf}}$. Note, that in configuration (b), the thickness is $2t$.

\section{Computational efficiency\label{sec:app_convergence}}

To evaluate the computation time of the algorithm proposed in this work, it was compared with a black-box trusted region framework algorithm implemented in the commercial package CST Microwave Studio \cite{cst}. The merit function for the optimization is $f_m= |S_{12}^\mathrm{re}-S_{12}^\mathrm{des}|^2$, as previously. By far the most significant contribution to the computation time comes from full wave simulations. Therefore, we compare here the number of full wave simulations $n_{fw}$ that is required until the respective optimization algorithm converges for both of these  methods. Since two full wave simulations are required to determine the near-field impedance matrix $\mathbf{Z_{nf}}$, the proposed model based optimization requires  $n_{fw}=2p_t$ when terminating at iteration $p_t$. The black-box optimization on the other hand requires $n_{fw}=p_t+1$. 

\begin{figure}[htbp]
	\includegraphics[width=.40\textwidth]{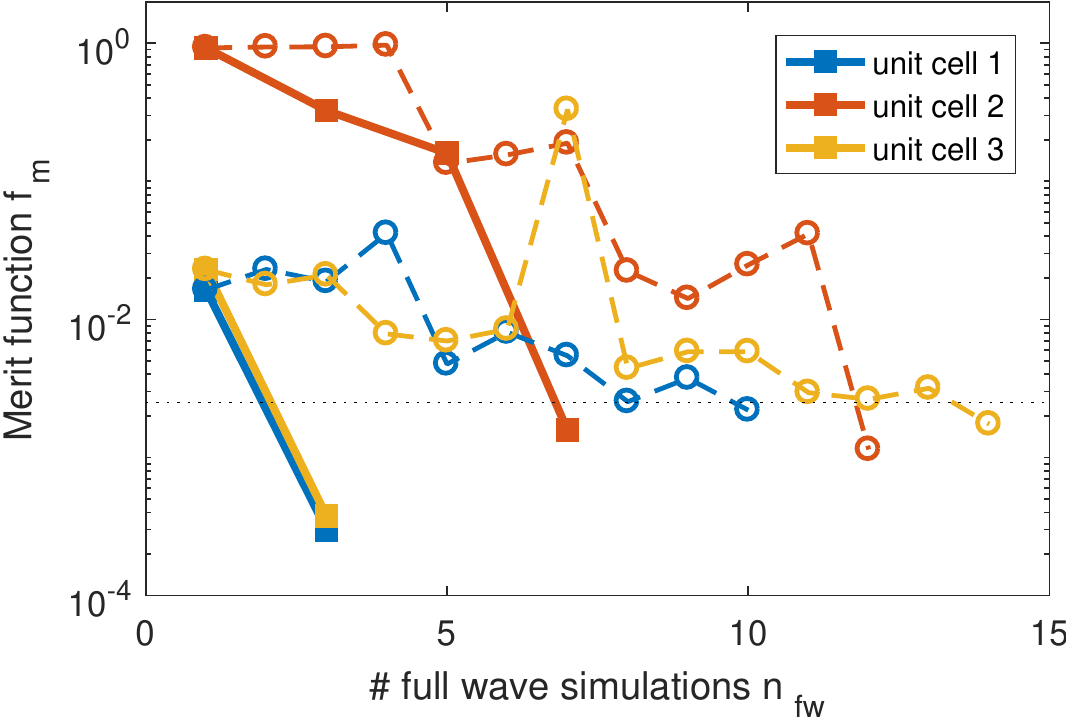}
	\caption{Comparison of the computation effort of the proposed model based optimization (solid lines with squares) and a black-box optimization algorithm (dashed lines with circles).  \label{fig:convergence_computation}}	
\end{figure}

In Figure~\ref{fig:convergence_computation}, the computational effort for the two algorithms is compared in terms of $n_{fw}$, for each of the three cells in our anomalously refracting metasurface. Our proposed optimization technique was on average 3.6 times faster than the black-box optimization. Additionally, the trust region optimization requires some prior knowledge to set the boundaries of the parameter space. Here, we chose $\pm 25\% $ on the initial parameter set. \\
\\

%\FloatBarrier

%\bibliography{Interlayer_coupling}

\begin{thebibliography}{41}%
\makeatletter
\providecommand \@ifxundefined [1]{%
 \@ifx{#1\undefined}
}%
\providecommand \@ifnum [1]{%
 \ifnum #1\expandafter \@firstoftwo
 \else \expandafter \@secondoftwo
 \fi
}%
\providecommand \@ifx [1]{%
 \ifx #1\expandafter \@firstoftwo
 \else \expandafter \@secondoftwo
 \fi
}%
\providecommand \natexlab [1]{#1}%
\providecommand \enquote  [1]{``#1''}%
\providecommand \bibnamefont  [1]{#1}%
\providecommand \bibfnamefont [1]{#1}%
\providecommand \citenamefont [1]{#1}%
\providecommand \href@noop [0]{\@secondoftwo}%
\providecommand \href [0]{\begingroup \@sanitize@url \@href}%
\providecommand \@href[1]{\@@startlink{#1}\@@href}%
\providecommand \@@href[1]{\endgroup#1\@@endlink}%
\providecommand \@sanitize@url [0]{\catcode `\\12\catcode `\$12\catcode
  `\&12\catcode `\#12\catcode `\^12\catcode `\_12\catcode `\%12\relax}%
\providecommand \@@startlink[1]{}%
\providecommand \@@endlink[0]{}%
\providecommand \url  [0]{\begingroup\@sanitize@url \@url }%
\providecommand \@url [1]{\endgroup\@href {#1}{\urlprefix }}%
\providecommand \urlprefix  [0]{URL }%
\providecommand \Eprint [0]{\href }%
\providecommand \doibase [0]{http://dx.doi.org/}%
\providecommand \selectlanguage [0]{\@gobble}%
\providecommand \bibinfo  [0]{\@secondoftwo}%
\providecommand \bibfield  [0]{\@secondoftwo}%
\providecommand \translation [1]{[#1]}%
\providecommand \BibitemOpen [0]{}%
\providecommand \bibitemStop [0]{}%
\providecommand \bibitemNoStop [0]{.\EOS\space}%
\providecommand \EOS [0]{\spacefactor3000\relax}%
\providecommand \BibitemShut  [1]{\csname bibitem#1\endcsname}%
\let\auto@bib@innerbib\@empty
%</preamble>
\bibitem [{\citenamefont {Yu}\ and\ \citenamefont
  {Capasso}(2014)}]{yuFlatOpticsDesigner2014}%
  \BibitemOpen
  \bibfield  {author} {\bibinfo {author} {\bibfnamefont {Nanfang}\ \bibnamefont
  {Yu}}\ and\ \bibinfo {author} {\bibfnamefont {Federico}\ \bibnamefont
  {Capasso}},\ }\bibfield  {title} {\enquote {\bibinfo {title} {Flat optics
  with designer metasurfaces},}\ }\href {\doibase 10.1038/nmat3839} {\bibfield
  {journal} {\bibinfo  {journal} {Nature Materials}\ }\textbf {\bibinfo
  {volume} {13}},\ \bibinfo {pages} {139--150} (\bibinfo {year}
  {2014})}\BibitemShut {NoStop}%
\bibitem [{\citenamefont {Ding}\ \emph {et~al.}(2018)\citenamefont {Ding},
  \citenamefont {Pors},\ and\ \citenamefont
  {Bozhevolnyi}}]{dingGradientMetasurfacesReview2018}%
  \BibitemOpen
  \bibfield  {author} {\bibinfo {author} {\bibfnamefont {Fei}\ \bibnamefont
  {Ding}}, \bibinfo {author} {\bibfnamefont {Anders}\ \bibnamefont {Pors}}, \
  and\ \bibinfo {author} {\bibfnamefont {Sergey~I}\ \bibnamefont
  {Bozhevolnyi}},\ }\bibfield  {title} {\enquote {\bibinfo {title} {Gradient
  metasurfaces: A review of fundamentals and applications},}\ }\href {\doibase
  10.1088/1361-6633/aa8732} {\bibfield  {journal} {\bibinfo  {journal} {Reports
  on Progress in Physics}\ }\textbf {\bibinfo {volume} {81}},\ \bibinfo {pages}
  {026401} (\bibinfo {year} {2018})}\BibitemShut {NoStop}%
\bibitem [{\citenamefont {Chen}\ \emph {et~al.}(2018)\citenamefont {Chen},
  \citenamefont {Kim}, \citenamefont {Wong},\ and\ \citenamefont
  {Eleftheriades}}]{chenHuygensMetasurfacesMicrowaves2018}%
  \BibitemOpen
  \bibfield  {author} {\bibinfo {author} {\bibfnamefont {Michael}\ \bibnamefont
  {Chen}}, \bibinfo {author} {\bibfnamefont {Minseok}\ \bibnamefont {Kim}},
  \bibinfo {author} {\bibfnamefont {Alex~M.H.}\ \bibnamefont {Wong}}, \ and\
  \bibinfo {author} {\bibfnamefont {George~V.}\ \bibnamefont {Eleftheriades}},\
  }\bibfield  {title} {\enquote {\bibinfo {title} {Huygens' metasurfaces from
  microwaves to optics: A review},}\ }\href {\doibase 10.1515/nanoph-2017-0117}
  {\bibfield  {journal} {\bibinfo  {journal} {Nanophotonics}\ }\textbf
  {\bibinfo {volume} {7}} (\bibinfo {year} {2018}),\
  10.1515/nanoph-2017-0117}\BibitemShut {NoStop}%
\bibitem [{\citenamefont {Pfeiffer}\ and\ \citenamefont
  {Grbic}(2013{\natexlab{a}})}]{pfeifferMetamaterialHuygensSurfaces2013a}%
  \BibitemOpen
  \bibfield  {author} {\bibinfo {author} {\bibfnamefont {Carl}\ \bibnamefont
  {Pfeiffer}}\ and\ \bibinfo {author} {\bibfnamefont {Anthony}\ \bibnamefont
  {Grbic}},\ }\bibfield  {title} {\enquote {\bibinfo {title} {Metamaterial
  {{Huygens}}' {{Surfaces}}: {{Tailoring Wave Fronts}} with {{Reflectionless
  Sheets}}},}\ }\href {\doibase 10.1103/PhysRevLett.110.197401} {\bibfield
  {journal} {\bibinfo  {journal} {Physical Review Letters}\ }\textbf {\bibinfo
  {volume} {110}},\ \bibinfo {pages} {197401} (\bibinfo {year}
  {2013}{\natexlab{a}})}\BibitemShut {NoStop}%
\bibitem [{\citenamefont {Monticone}\ \emph {et~al.}(2013)\citenamefont
  {Monticone}, \citenamefont {Estakhri},\ and\ \citenamefont
  {Al\`u}}]{monticoneFullControlNanoscale2013}%
  \BibitemOpen
  \bibfield  {author} {\bibinfo {author} {\bibfnamefont {Francesco}\
  \bibnamefont {Monticone}}, \bibinfo {author} {\bibfnamefont
  {Nasim~Mohammadi}\ \bibnamefont {Estakhri}}, \ and\ \bibinfo {author}
  {\bibfnamefont {Andrea}\ \bibnamefont {Al\`u}},\ }\bibfield  {title}
  {\enquote {\bibinfo {title} {Full {{Control}} of {{Nanoscale Optical
  Transmission}} with a {{Composite Metascreen}}},}\ }\href {\doibase
  10.1103/PhysRevLett.110.203903} {\bibfield  {journal} {\bibinfo  {journal}
  {Physical Review Letters}\ }\textbf {\bibinfo {volume} {110}},\ \bibinfo
  {pages} {203903} (\bibinfo {year} {2013})}\BibitemShut {NoStop}%
\bibitem [{\citenamefont {Decker}\ \emph {et~al.}(2015)\citenamefont {Decker},
  \citenamefont {Staude}, \citenamefont {Falkner}, \citenamefont {Dominguez},
  \citenamefont {Neshev}, \citenamefont {Brener}, \citenamefont {Pertsch},\
  and\ \citenamefont {Kivshar}}]{deckerHighEfficiencyDielectricHuygens2015}%
  \BibitemOpen
  \bibfield  {author} {\bibinfo {author} {\bibfnamefont {Manuel}\ \bibnamefont
  {Decker}}, \bibinfo {author} {\bibfnamefont {Isabelle}\ \bibnamefont
  {Staude}}, \bibinfo {author} {\bibfnamefont {Matthias}\ \bibnamefont
  {Falkner}}, \bibinfo {author} {\bibfnamefont {Jason}\ \bibnamefont
  {Dominguez}}, \bibinfo {author} {\bibfnamefont {Dragomir~N.}\ \bibnamefont
  {Neshev}}, \bibinfo {author} {\bibfnamefont {Igal}\ \bibnamefont {Brener}},
  \bibinfo {author} {\bibfnamefont {Thomas}\ \bibnamefont {Pertsch}}, \ and\
  \bibinfo {author} {\bibfnamefont {Yuri~S.}\ \bibnamefont {Kivshar}},\
  }\bibfield  {title} {\enquote {\bibinfo {title} {High-{{Efficiency Dielectric
  Huygens}}' {{Surfaces}}},}\ }\href {\doibase 10.1002/adom.201400584}
  {\bibfield  {journal} {\bibinfo  {journal} {Advanced Optical Materials}\
  }\textbf {\bibinfo {volume} {3}},\ \bibinfo {pages} {813--820} (\bibinfo
  {year} {2015})}\BibitemShut {NoStop}%
\bibitem [{\citenamefont {Epstein}\ and\ \citenamefont
  {Eleftheriades}(2016{\natexlab{a}})}]{epsteinHuygensMetasurfacesEquivalence2016}%
  \BibitemOpen
  \bibfield  {author} {\bibinfo {author} {\bibfnamefont {Ariel}\ \bibnamefont
  {Epstein}}\ and\ \bibinfo {author} {\bibfnamefont {George~V.}\ \bibnamefont
  {Eleftheriades}},\ }\bibfield  {title} {\enquote {\bibinfo {title} {Huygens'
  metasurfaces via the equivalence principle: Design and applications},}\
  }\href@noop {} {\bibfield  {journal} {\bibinfo  {journal} {Journal of the
  Optical Society of America B}\ }\textbf {\bibinfo {volume} {33}},\ \bibinfo
  {pages} {A31--A50} (\bibinfo {year} {2016}{\natexlab{a}})}\BibitemShut
  {NoStop}%
\bibitem [{\citenamefont {Pfeiffer}\ and\ \citenamefont
  {Grbic}(2014)}]{pfeifferBianisotropicMetasurfacesOptimal2014}%
  \BibitemOpen
  \bibfield  {author} {\bibinfo {author} {\bibfnamefont {Carl}\ \bibnamefont
  {Pfeiffer}}\ and\ \bibinfo {author} {\bibfnamefont {Anthony}\ \bibnamefont
  {Grbic}},\ }\bibfield  {title} {\enquote {\bibinfo {title} {Bianisotropic
  {{Metasurfaces}} for {{Optimal Polarization Control}}: {{Analysis}} and
  {{Synthesis}}},}\ }\href {\doibase 10.1103/PhysRevApplied.2.044011}
  {\bibfield  {journal} {\bibinfo  {journal} {Physical Review Applied}\
  }\textbf {\bibinfo {volume} {2}} (\bibinfo {year} {2014}),\
  10.1103/PhysRevApplied.2.044011}\BibitemShut {NoStop}%
\bibitem [{\citenamefont {Wong}\ \emph {et~al.}(2016)\citenamefont {Wong},
  \citenamefont {Epstein},\ and\ \citenamefont
  {Eleftheriades}}]{wongReflectionlessWideAngleRefracting2016}%
  \BibitemOpen
  \bibfield  {author} {\bibinfo {author} {\bibfnamefont {Joseph P.~S.}\
  \bibnamefont {Wong}}, \bibinfo {author} {\bibfnamefont {Ariel}\ \bibnamefont
  {Epstein}}, \ and\ \bibinfo {author} {\bibfnamefont {George~V.}\ \bibnamefont
  {Eleftheriades}},\ }\bibfield  {title} {\enquote {\bibinfo {title}
  {Reflectionless {{Wide}}-{{Angle Refracting Metasurfaces}}},}\ }\href
  {\doibase 10.1109/LAWP.2015.2505629} {\bibfield  {journal} {\bibinfo
  {journal} {IEEE Antennas and Wireless Propagation Letters}\ }\textbf
  {\bibinfo {volume} {15}},\ \bibinfo {pages} {1293--1296} (\bibinfo {year}
  {2016})}\BibitemShut {NoStop}%
\bibitem [{\citenamefont {Cole}\ \emph {et~al.}(2018)\citenamefont {Cole},
  \citenamefont {Lamprianidis}, \citenamefont {Shadrivov},\ and\ \citenamefont
  {Powell}}]{ColeRefractionefficiencyHuygens2018}%
  \BibitemOpen
  \bibfield  {author} {\bibinfo {author} {\bibfnamefont {Michael~A.}\
  \bibnamefont {Cole}}, \bibinfo {author} {\bibfnamefont {Aristeidis}\
  \bibnamefont {Lamprianidis}}, \bibinfo {author} {\bibfnamefont {Ilya~V.}\
  \bibnamefont {Shadrivov}}, \ and\ \bibinfo {author} {\bibfnamefont
  {David~A.}\ \bibnamefont {Powell}},\ }\bibfield  {title} {\enquote {\bibinfo
  {title} {Refraction efficiency of {{Huygens}}' and bianisotropic terahertz
  metasurfaces},}\ }\href@noop {} {\bibfield  {journal} {\bibinfo  {journal}
  {arXiv:1812.04725 [physics]}\ } (\bibinfo {year} {2018})}\BibitemShut
  {NoStop}%
\bibitem [{\citenamefont {Fathnan}\ and\ \citenamefont
  {Powell}(2018)}]{fathnanBandwidthSizeLimits2018}%
  \BibitemOpen
  \bibfield  {author} {\bibinfo {author} {\bibfnamefont {Ashif~A.}\
  \bibnamefont {Fathnan}}\ and\ \bibinfo {author} {\bibfnamefont {David~A.}\
  \bibnamefont {Powell}},\ }\bibfield  {title} {\enquote {\bibinfo {title}
  {Bandwidth and size limits of achromatic printed-circuit metasurfaces},}\
  }\href {\doibase 10.1364/OE.26.029440} {\bibfield  {journal} {\bibinfo
  {journal} {Optics Express}\ }\textbf {\bibinfo {volume} {26}},\ \bibinfo
  {pages} {29440} (\bibinfo {year} {2018})}\BibitemShut {NoStop}%
\bibitem [{\citenamefont {Wong}\ \emph {et~al.}(2014)\citenamefont {Wong},
  \citenamefont {Selvanayagam},\ and\ \citenamefont
  {Eleftheriades}}]{wongDesignUnitCells2014}%
  \BibitemOpen
  \bibfield  {author} {\bibinfo {author} {\bibfnamefont {Joseph P.~S.}\
  \bibnamefont {Wong}}, \bibinfo {author} {\bibfnamefont {Michael}\
  \bibnamefont {Selvanayagam}}, \ and\ \bibinfo {author} {\bibfnamefont
  {George~V.}\ \bibnamefont {Eleftheriades}},\ }\bibfield  {title} {\enquote
  {\bibinfo {title} {Design of unit cells and demonstration of methods for
  synthesizing {{Huygens}} metasurfaces},}\ }\href {\doibase
  10.1016/j.photonics.2014.07.001} {\bibfield  {journal} {\bibinfo  {journal}
  {Photonics and Nanostructures - Fundamentals and Applications}\ }\bibinfo
  {series} {Metamaterials-2013 Congress},\ \textbf {\bibinfo {volume} {12}},\
  \bibinfo {pages} {360--375} (\bibinfo {year} {2014})}\BibitemShut {NoStop}%
\bibitem [{\citenamefont {Lavigne}\ \emph {et~al.}(2018)\citenamefont
  {Lavigne}, \citenamefont {Achouri}, \citenamefont {Asadchy}, \citenamefont
  {Tretyakov},\ and\ \citenamefont
  {Caloz}}]{lavigneSusceptibilityDerivationExperimental2018}%
  \BibitemOpen
  \bibfield  {author} {\bibinfo {author} {\bibfnamefont {G.}~\bibnamefont
  {Lavigne}}, \bibinfo {author} {\bibfnamefont {K.}~\bibnamefont {Achouri}},
  \bibinfo {author} {\bibfnamefont {V.~S.}\ \bibnamefont {Asadchy}}, \bibinfo
  {author} {\bibfnamefont {S.~A.}\ \bibnamefont {Tretyakov}}, \ and\ \bibinfo
  {author} {\bibfnamefont {C.}~\bibnamefont {Caloz}},\ }\bibfield  {title}
  {\enquote {\bibinfo {title} {Susceptibility {{Derivation}} and {{Experimental
  Demonstration}} of {{Refracting Metasurfaces Without Spurious
  Diffraction}}},}\ }\href {\doibase 10.1109/TAP.2018.2793958} {\bibfield
  {journal} {\bibinfo  {journal} {IEEE Transactions on Antennas and
  Propagation}\ }\textbf {\bibinfo {volume} {66}},\ \bibinfo {pages}
  {1321--1330} (\bibinfo {year} {2018})}\BibitemShut {NoStop}%
\bibitem [{\citenamefont {Capolino}\ \emph {et~al.}(2013)\citenamefont
  {Capolino}, \citenamefont {Vallecchi},\ and\ \citenamefont
  {Albani}}]{capolinoEquivalentTransmissionLine2013}%
  \BibitemOpen
  \bibfield  {author} {\bibinfo {author} {\bibfnamefont {Filippo}\ \bibnamefont
  {Capolino}}, \bibinfo {author} {\bibfnamefont {Andrea}\ \bibnamefont
  {Vallecchi}}, \ and\ \bibinfo {author} {\bibfnamefont {Matteo}\ \bibnamefont
  {Albani}},\ }\bibfield  {title} {\enquote {\bibinfo {title} {Equivalent
  {{Transmission Line Model With}} a {{Lumped X}}-{{Circuit}} for a {{Metalayer
  Made}} of {{Pairs}} of {{Planar Conductors}}},}\ }\href {\doibase
  10.1109/TAP.2012.2225013} {\bibfield  {journal} {\bibinfo  {journal} {IEEE
  Transactions on Antennas and Propagation}\ }\textbf {\bibinfo {volume}
  {61}},\ \bibinfo {pages} {852--861} (\bibinfo {year} {2013})}\BibitemShut
  {NoStop}%
\bibitem [{\citenamefont {Rabinovich}\ and\ \citenamefont
  {Epstein}(2018)}]{rabinovichAnalyticalDesignPrintedCircuitBoard2018}%
  \BibitemOpen
  \bibfield  {author} {\bibinfo {author} {\bibfnamefont {Oshri}\ \bibnamefont
  {Rabinovich}}\ and\ \bibinfo {author} {\bibfnamefont {Ariel}\ \bibnamefont
  {Epstein}},\ }\bibfield  {title} {\enquote {\bibinfo {title} {Analytical
  {{Design}} of {{Printed}}-{{Circuit}}-{{Board}} ({{PCB}}) {{Metagratings}}
  for {{Perfect Anomalous Reflection}}},}\ }\href {\doibase
  10.1109/TAP.2018.2836379} {\bibfield  {journal} {\bibinfo  {journal} {IEEE
  Transactions on Antennas and Propagation}\ ,\ \bibinfo {pages} {0018--926X,
  1558--2221}} (\bibinfo {year} {2018})}\BibitemShut {NoStop}%
\bibitem [{\citenamefont {Rappaport}\ \emph {et~al.}()\citenamefont
  {Rappaport}, \citenamefont {Xing}, \citenamefont {MacCartney}, \citenamefont
  {Molisch}, \citenamefont {Mellios},\ and\ \citenamefont
  {Zhang}}]{rappaportOverviewMillimeterWave2017}%
  \BibitemOpen
  \bibfield  {author} {\bibinfo {author} {\bibfnamefont {Theodore~S.}\
  \bibnamefont {Rappaport}}, \bibinfo {author} {\bibfnamefont {Yunchou}\
  \bibnamefont {Xing}}, \bibinfo {author} {\bibfnamefont {George~R.}\
  \bibnamefont {MacCartney}}, \bibinfo {author} {\bibfnamefont {Andreas~F.}\
  \bibnamefont {Molisch}}, \bibinfo {author} {\bibfnamefont {Evangelos}\
  \bibnamefont {Mellios}}, \ and\ \bibinfo {author} {\bibfnamefont {Jianhua}\
  \bibnamefont {Zhang}},\ }\bibfield  {title} {\enquote {\bibinfo {title}
  {Overview of {{Millimeter Wave Communications}} for {{Fifth}}-{{Generation}}
  ({{5G}}) {{Wireless Networks}}—{{With}} a {{Focus}} on {{Propagation
  Models}}},}\ }\href {\doibase 10.1109/TAP.2017.2734243} {\ \textbf {\bibinfo
  {volume} {65}},\ \bibinfo {pages} {6213--6230}}\BibitemShut {NoStop}%
\bibitem [{\citenamefont {Patole}\ \emph {et~al.}()\citenamefont {Patole},
  \citenamefont {Torlak}, \citenamefont {Wang},\ and\ \citenamefont
  {Ali}}]{patoleAutomotiveRadarsReview2017}%
  \BibitemOpen
  \bibfield  {author} {\bibinfo {author} {\bibfnamefont {S.~M.}\ \bibnamefont
  {Patole}}, \bibinfo {author} {\bibfnamefont {M.}~\bibnamefont {Torlak}},
  \bibinfo {author} {\bibfnamefont {D.}~\bibnamefont {Wang}}, \ and\ \bibinfo
  {author} {\bibfnamefont {M.}~\bibnamefont {Ali}},\ }\bibfield  {title}
  {\enquote {\bibinfo {title} {Automotive radars: {{A}} review of signal
  processing techniques},}\ }\href {\doibase 10.1109/MSP.2016.2628914} {\
  \textbf {\bibinfo {volume} {34}},\ \bibinfo {pages} {22--35}}\BibitemShut
  {NoStop}%
\bibitem [{\citenamefont {Hasch}\ \emph {et~al.}()\citenamefont {Hasch},
  \citenamefont {Topak}, \citenamefont {Schnabel}, \citenamefont {Zwick},
  \citenamefont {Weigel},\ and\ \citenamefont
  {Waldschmidt}}]{haschMillimeterWaveTechnologyAutomotive2012}%
  \BibitemOpen
  \bibfield  {author} {\bibinfo {author} {\bibfnamefont {J.}~\bibnamefont
  {Hasch}}, \bibinfo {author} {\bibfnamefont {E.}~\bibnamefont {Topak}},
  \bibinfo {author} {\bibfnamefont {R.}~\bibnamefont {Schnabel}}, \bibinfo
  {author} {\bibfnamefont {T.}~\bibnamefont {Zwick}}, \bibinfo {author}
  {\bibfnamefont {R.}~\bibnamefont {Weigel}}, \ and\ \bibinfo {author}
  {\bibfnamefont {C.}~\bibnamefont {Waldschmidt}},\ }\bibfield  {title}
  {\enquote {\bibinfo {title} {Millimeter-{{Wave Technology}} for {{Automotive
  Radar Sensors}} in the 77 {{GHz Frequency Band}}},}\ }\href {\doibase
  10.1109/TMTT.2011.2178427} {\ \textbf {\bibinfo {volume} {60}},\ \bibinfo
  {pages} {845--860}}\BibitemShut {NoStop}%
\bibitem [{\citenamefont {Pfeiffer}\ and\ \citenamefont
  {Grbic}(2013{\natexlab{b}})}]{pfeifferMillimeterWaveTransmitarraysWavefront2013}%
  \BibitemOpen
  \bibfield  {author} {\bibinfo {author} {\bibfnamefont {C.}~\bibnamefont
  {Pfeiffer}}\ and\ \bibinfo {author} {\bibfnamefont {A.}~\bibnamefont
  {Grbic}},\ }\bibfield  {title} {\enquote {\bibinfo {title} {Millimeter-{{Wave
  Transmitarrays}} for {{Wavefront}} and {{Polarization Control}}},}\ }\href
  {\doibase 10.1109/TMTT.2013.2287173} {\bibfield  {journal} {\bibinfo
  {journal} {IEEE Transactions on Microwave Theory and Techniques}\ }\textbf
  {\bibinfo {volume} {61}},\ \bibinfo {pages} {4407--4417} (\bibinfo {year}
  {2013}{\natexlab{b}})}\BibitemShut {NoStop}%
\bibitem [{\citenamefont {Jiang}\ \emph {et~al.}()\citenamefont {Jiang},
  \citenamefont {Kang}, \citenamefont {Hong},\ and\ \citenamefont
  {Werner}}]{jiangHighlyEfficientBroadband2018}%
  \BibitemOpen
  \bibfield  {author} {\bibinfo {author} {\bibfnamefont {Zhi~Hao}\ \bibnamefont
  {Jiang}}, \bibinfo {author} {\bibfnamefont {Lei}\ \bibnamefont {Kang}},
  \bibinfo {author} {\bibfnamefont {Wei}\ \bibnamefont {Hong}}, \ and\ \bibinfo
  {author} {\bibfnamefont {Douglas~H.}\ \bibnamefont {Werner}},\ }\bibfield
  {title} {\enquote {\bibinfo {title} {Highly {{Efficient Broadband Multiplexed
  Millimeter}}-{{Wave Vortices}} from {{Metasurface}}-{{Enabled
  Transmit}}-{{Arrays}} of {{Subwavelength Thickness}}},}\ }\href {\doibase
  10.1103/PhysRevApplied.9.064009} {\ \textbf {\bibinfo {volume} {9}},\
  10.1103/PhysRevApplied.9.064009}\BibitemShut {NoStop}%
\bibitem [{\citenamefont {Epstein}\ and\ \citenamefont
  {Eleftheriades}(2016{\natexlab{b}})}]{epsteinArbitraryPowerConservingField2016}%
  \BibitemOpen
  \bibfield  {author} {\bibinfo {author} {\bibfnamefont {Ariel}\ \bibnamefont
  {Epstein}}\ and\ \bibinfo {author} {\bibfnamefont {George~V.}\ \bibnamefont
  {Eleftheriades}},\ }\bibfield  {title} {\enquote {\bibinfo {title} {Arbitrary
  {{Power}}-{{Conserving Field Transformations With Passive Lossless
  Omega}}-{{Type Bianisotropic Metasurfaces}}},}\ }\href {\doibase
  10.1109/TAP.2016.2588495} {\bibfield  {journal} {\bibinfo  {journal} {IEEE
  Transactions on Antennas and Propagation}\ }\textbf {\bibinfo {volume}
  {64}},\ \bibinfo {pages} {3880--3895} (\bibinfo {year}
  {2016}{\natexlab{b}})}\BibitemShut {NoStop}%
\bibitem [{\citenamefont {Epstein}\ and\ \citenamefont
  {Eleftheriades}()}]{epsteinSynthesisPassiveLossless2016}%
  \BibitemOpen
  \bibfield  {author} {\bibinfo {author} {\bibfnamefont {Ariel}\ \bibnamefont
  {Epstein}}\ and\ \bibinfo {author} {\bibfnamefont {George~V.}\ \bibnamefont
  {Eleftheriades}},\ }\bibfield  {title} {\enquote {\bibinfo {title} {Synthesis
  of {{Passive Lossless Metasurfaces Using Auxiliary Fields}} for
  {{Reflectionless Beam Splitting}} and {{Perfect Reflection}}},}\ }\href
  {\doibase 10.1103/PhysRevLett.117.256103} {\ \textbf {\bibinfo {volume}
  {117}},\ 10.1103/PhysRevLett.117.256103}\BibitemShut {NoStop}%
\bibitem [{\citenamefont {Syms}\ \emph {et~al.}(2005)\citenamefont {Syms},
  \citenamefont {Shamonina}, \citenamefont {Kalinin},\ and\ \citenamefont
  {Solymar}}]{Symstheorymetamaterialsbased2005}%
  \BibitemOpen
  \bibfield  {author} {\bibinfo {author} {\bibfnamefont {R.~R.~A.}\
  \bibnamefont {Syms}}, \bibinfo {author} {\bibfnamefont {Ekaterina}\
  \bibnamefont {Shamonina}}, \bibinfo {author} {\bibfnamefont {V.}~\bibnamefont
  {Kalinin}}, \ and\ \bibinfo {author} {\bibfnamefont {L.}~\bibnamefont
  {Solymar}},\ }\bibfield  {title} {\enquote {\bibinfo {title} {A theory of
  metamaterials based on periodically loaded transmission lines:
  {{Interaction}} between magnetoinductive and electromagnetic waves},}\ }\href
  {\doibase 10.1063/1.1850182} {\bibfield  {journal} {\bibinfo  {journal}
  {Journal of Applied Physics}\ }\textbf {\bibinfo {volume} {97}},\ \bibinfo
  {pages} {064909--064909} (\bibinfo {year} {2005})}\BibitemShut {NoStop}%
\bibitem [{\citenamefont {Powell}\ \emph {et~al.}(2010)\citenamefont {Powell},
  \citenamefont {Lapine}, \citenamefont {Gorkunov}, \citenamefont {Shadrivov},\
  and\ \citenamefont {Kivshar}}]{powellMetamaterialTuningManipulation2010}%
  \BibitemOpen
  \bibfield  {author} {\bibinfo {author} {\bibfnamefont {David~A.}\
  \bibnamefont {Powell}}, \bibinfo {author} {\bibfnamefont {Mikhail}\
  \bibnamefont {Lapine}}, \bibinfo {author} {\bibfnamefont {Maxim~V.}\
  \bibnamefont {Gorkunov}}, \bibinfo {author} {\bibfnamefont {Ilya~V.}\
  \bibnamefont {Shadrivov}}, \ and\ \bibinfo {author} {\bibfnamefont {Yuri~S.}\
  \bibnamefont {Kivshar}},\ }\bibfield  {title} {\enquote {\bibinfo {title}
  {Metamaterial tuning by manipulation of near-field interaction},}\ }\href
  {\doibase 10.1103/PhysRevB.82.155128} {\bibfield  {journal} {\bibinfo
  {journal} {Physical Review B}\ }\textbf {\bibinfo {volume} {82}} (\bibinfo
  {year} {2010}),\ 10.1103/PhysRevB.82.155128}\BibitemShut {NoStop}%
\bibitem [{\citenamefont {Liu}\ \emph {et~al.}(2007)\citenamefont {Liu},
  \citenamefont {Genov}, \citenamefont {Wu}, \citenamefont {Liu}, \citenamefont
  {Liu}, \citenamefont {Sun}, \citenamefont {Zhu},\ and\ \citenamefont
  {Zhang}}]{LiuMagneticplasmonhybridization2007}%
  \BibitemOpen
  \bibfield  {author} {\bibinfo {author} {\bibfnamefont {H.}~\bibnamefont
  {Liu}}, \bibinfo {author} {\bibfnamefont {D.}~\bibnamefont {Genov}}, \bibinfo
  {author} {\bibfnamefont {D.}~\bibnamefont {Wu}}, \bibinfo {author}
  {\bibfnamefont {Y.}~\bibnamefont {Liu}}, \bibinfo {author} {\bibfnamefont
  {Z.}~\bibnamefont {Liu}}, \bibinfo {author} {\bibfnamefont {C.}~\bibnamefont
  {Sun}}, \bibinfo {author} {\bibfnamefont {S.}~\bibnamefont {Zhu}}, \ and\
  \bibinfo {author} {\bibfnamefont {X.}~\bibnamefont {Zhang}},\ }\bibfield
  {title} {\enquote {\bibinfo {title} {Magnetic plasmon hybridization and
  optical activity at optical frequencies in metallic nanostructures},}\ }\href
  {\doibase 10.1103/PhysRevB.76.073101} {\bibfield  {journal} {\bibinfo
  {journal} {Physical Review B}\ }\textbf {\bibinfo {volume} {76}},\ \bibinfo
  {pages} {073101--073101} (\bibinfo {year} {2007})}\BibitemShut {NoStop}%
\bibitem [{\citenamefont {Liu}\ \emph {et~al.}(2012)\citenamefont {Liu},
  \citenamefont {Powell}, \citenamefont {Shadrivov},\ and\ \citenamefont
  {Kivshar}}]{liuOpticalActivityCoupling2012}%
  \BibitemOpen
  \bibfield  {author} {\bibinfo {author} {\bibfnamefont {Mingkai}\ \bibnamefont
  {Liu}}, \bibinfo {author} {\bibfnamefont {David~A.}\ \bibnamefont {Powell}},
  \bibinfo {author} {\bibfnamefont {Ilya~V.}\ \bibnamefont {Shadrivov}}, \ and\
  \bibinfo {author} {\bibfnamefont {Yuri~S.}\ \bibnamefont {Kivshar}},\
  }\bibfield  {title} {\enquote {\bibinfo {title} {Optical activity and
  coupling in twisted dimer meta-atoms},}\ }\href {\doibase 10.1063/1.3694269}
  {\bibfield  {journal} {\bibinfo  {journal} {Applied Physics Letters}\
  }\textbf {\bibinfo {volume} {100}},\ \bibinfo {pages} {111114} (\bibinfo
  {year} {2012})}\BibitemShut {NoStop}%
\bibitem [{\citenamefont {Chalabi}\ \emph {et~al.}(2017)\citenamefont
  {Chalabi}, \citenamefont {Ra'di}, \citenamefont {Sounas},\ and\ \citenamefont
  {Al\`u}}]{chalabiEfficientAnomalousReflection2017}%
  \BibitemOpen
  \bibfield  {author} {\bibinfo {author} {\bibfnamefont {H.}~\bibnamefont
  {Chalabi}}, \bibinfo {author} {\bibfnamefont {Y.}~\bibnamefont {Ra'di}},
  \bibinfo {author} {\bibfnamefont {D.~L.}\ \bibnamefont {Sounas}}, \ and\
  \bibinfo {author} {\bibfnamefont {A.}~\bibnamefont {Al\`u}},\ }\bibfield
  {title} {\enquote {\bibinfo {title} {Efficient anomalous reflection through
  near-field interactions in metasurfaces},}\ }\href {\doibase
  10.1103/PhysRevB.96.075432} {\bibfield  {journal} {\bibinfo  {journal}
  {Physical Review B}\ }\textbf {\bibinfo {volume} {96}} (\bibinfo {year}
  {2017}),\ 10.1103/PhysRevB.96.075432}\BibitemShut {NoStop}%
\bibitem [{\citenamefont {Ra'di}\ \emph {et~al.}(2017)\citenamefont {Ra'di},
  \citenamefont {Sounas},\ and\ \citenamefont
  {Al\`u}}]{radiMetagratingsLimitsGraded2017}%
  \BibitemOpen
  \bibfield  {author} {\bibinfo {author} {\bibfnamefont {Younes}\ \bibnamefont
  {Ra'di}}, \bibinfo {author} {\bibfnamefont {Dimitrios~L.}\ \bibnamefont
  {Sounas}}, \ and\ \bibinfo {author} {\bibfnamefont {Andrea}\ \bibnamefont
  {Al\`u}},\ }\bibfield  {title} {\enquote {\bibinfo {title} {Metagratings:
  {{Beyond}} the {{Limits}} of {{Graded Metasurfaces}} for {{Wave Front
  Control}}},}\ }\href {\doibase 10.1103/PhysRevLett.119.067404} {\bibfield
  {journal} {\bibinfo  {journal} {Physical Review Letters}\ }\textbf {\bibinfo
  {volume} {119}} (\bibinfo {year} {2017}),\
  10.1103/PhysRevLett.119.067404}\BibitemShut {NoStop}%
\bibitem [{\citenamefont {Popov}\ \emph {et~al.}(2018)\citenamefont {Popov},
  \citenamefont {Boust},\ and\ \citenamefont
  {Burokur}}]{popovControllingDiffractionPatterns2018}%
  \BibitemOpen
  \bibfield  {author} {\bibinfo {author} {\bibfnamefont {Vladislav}\
  \bibnamefont {Popov}}, \bibinfo {author} {\bibfnamefont {Fabrice}\
  \bibnamefont {Boust}}, \ and\ \bibinfo {author} {\bibfnamefont {Shah~Nawaz}\
  \bibnamefont {Burokur}},\ }\bibfield  {title} {\enquote {\bibinfo {title}
  {Controlling {{Diffraction Patterns}} with {{Metagratings}}},}\ }\href
  {\doibase 10.1103/PhysRevApplied.10.011002} {\bibfield  {journal} {\bibinfo
  {journal} {Physical Review Applied}\ }\textbf {\bibinfo {volume} {10}}
  (\bibinfo {year} {2018}),\ 10.1103/PhysRevApplied.10.011002}\BibitemShut
  {NoStop}%
\bibitem [{\citenamefont {Epstein}\ and\ \citenamefont
  {Rabinovich}(2018)}]{epsteinPerfectAnomalousRefraction2018}%
  \BibitemOpen
  \bibfield  {author} {\bibinfo {author} {\bibfnamefont {Ariel}\ \bibnamefont
  {Epstein}}\ and\ \bibinfo {author} {\bibfnamefont {Oshri}\ \bibnamefont
  {Rabinovich}},\ }\bibfield  {title} {\enquote {\bibinfo {title} {Perfect
  {{Anomalous Refraction}} with {{Metagratings}}},}\ }\href@noop {} {\bibfield
  {journal} {\bibinfo  {journal} {arXiv:1804.02362 [physics]}\ } (\bibinfo
  {year} {2018})}\BibitemShut {NoStop}%
\bibitem [{\citenamefont {Deshpande}\ \emph {et~al.}(2018)\citenamefont
  {Deshpande}, \citenamefont {Zenin}, \citenamefont {Ding}, \citenamefont
  {Mortensen},\ and\ \citenamefont
  {Bozhevolnyi}}]{deshpandeDirectCharacterizationNearField2018}%
  \BibitemOpen
  \bibfield  {author} {\bibinfo {author} {\bibfnamefont {Rucha}\ \bibnamefont
  {Deshpande}}, \bibinfo {author} {\bibfnamefont {Vladimir~A.}\ \bibnamefont
  {Zenin}}, \bibinfo {author} {\bibfnamefont {Fei}\ \bibnamefont {Ding}},
  \bibinfo {author} {\bibfnamefont {N.~Asger}\ \bibnamefont {Mortensen}}, \
  and\ \bibinfo {author} {\bibfnamefont {Sergey~I.}\ \bibnamefont
  {Bozhevolnyi}},\ }\bibfield  {title} {\enquote {\bibinfo {title} {Direct
  {{Characterization}} of {{Near}}-{{Field Coupling}} in {{Gap
  Plasmon}}-{{Based Metasurfaces}}},}\ }\href {\doibase
  10.1021/acs.nanolett.8b02393} {\bibfield  {journal} {\bibinfo  {journal}
  {Nano Letters}\ } (\bibinfo {year} {2018}),\
  10.1021/acs.nanolett.8b02393}\BibitemShut {NoStop}%
\bibitem [{\citenamefont {Sharac}\ \emph {et~al.}()\citenamefont {Sharac},
  \citenamefont {Giles}, \citenamefont {Perkins}, \citenamefont {Tischler},
  \citenamefont {Bezares}, \citenamefont {Prokes}, \citenamefont {Folland},
  \citenamefont {Glembocki},\ and\ \citenamefont
  {Caldwell}}]{sharacImplementationPlasmonicBand2018}%
  \BibitemOpen
  \bibfield  {author} {\bibinfo {author} {\bibfnamefont {Nicholas}\
  \bibnamefont {Sharac}}, \bibinfo {author} {\bibfnamefont {Alexander~J.}\
  \bibnamefont {Giles}}, \bibinfo {author} {\bibfnamefont {Keith}\ \bibnamefont
  {Perkins}}, \bibinfo {author} {\bibfnamefont {Joseph}\ \bibnamefont
  {Tischler}}, \bibinfo {author} {\bibfnamefont {Francisco}\ \bibnamefont
  {Bezares}}, \bibinfo {author} {\bibfnamefont {Sharka~M.}\ \bibnamefont
  {Prokes}}, \bibinfo {author} {\bibfnamefont {Thomas~G.}\ \bibnamefont
  {Folland}}, \bibinfo {author} {\bibfnamefont {Orest~J.}\ \bibnamefont
  {Glembocki}}, \ and\ \bibinfo {author} {\bibfnamefont {Joshua~D.}\
  \bibnamefont {Caldwell}},\ }\bibfield  {title} {\enquote {\bibinfo {title}
  {Implementation of plasmonic band structure to understand polariton
  hybridization within metamaterials},}\ }\href {\doibase 10.1364/OE.26.029363}
  {\ \textbf {\bibinfo {volume} {26}},\ \bibinfo {pages} {29363}}\BibitemShut
  {NoStop}%
\bibitem [{\citenamefont
  {Tretyakov}(2003)}]{tretyakovAnalyticalMethodsApplied2003}%
  \BibitemOpen
  \bibfield  {author} {\bibinfo {author} {\bibfnamefont {Sergei}\ \bibnamefont
  {Tretyakov}},\ }\href@noop {} {\emph {\bibinfo {title} {Analytical
  {{Methods}} in {{Applied Electrodynamics}}}}}\ (\bibinfo  {publisher}
  {{Artech House}},\ \bibinfo {year} {2003})\BibitemShut {NoStop}%
\bibitem [{\citenamefont {{Abdo-Sanchez}}\ \emph {et~al.}(2018)\citenamefont
  {{Abdo-Sanchez}}, \citenamefont {Chen}, \citenamefont {Epstein},\ and\
  \citenamefont {Eleftheriades}}]{abdo-sanchezLeakyWaveAntennaControlled2018}%
  \BibitemOpen
  \bibfield  {author} {\bibinfo {author} {\bibfnamefont {Elena}\ \bibnamefont
  {{Abdo-Sanchez}}}, \bibinfo {author} {\bibfnamefont {Michael}\ \bibnamefont
  {Chen}}, \bibinfo {author} {\bibfnamefont {Ariel}\ \bibnamefont {Epstein}}, \
  and\ \bibinfo {author} {\bibfnamefont {George~V.}\ \bibnamefont
  {Eleftheriades}},\ }\bibfield  {title} {\enquote {\bibinfo {title} {A
  {{Leaky}}-{{Wave Antenna With Controlled Radiation Using}} a {{Bianisotropic
  Huygens}}' {{Metasurface}}},}\ }\href {\doibase 10.1109/TAP.2018.2878082}
  {\bibfield  {journal} {\bibinfo  {journal} {IEEE Transactions on Antennas and
  Propagation}\ ,\ \bibinfo {pages} {0018--926X, 1558--2221}} (\bibinfo {year}
  {2018})}\BibitemShut {NoStop}%
\bibitem [{\citenamefont {Liu}\ \emph {et~al.}(2016)\citenamefont {Liu},
  \citenamefont {Bai}, \citenamefont {Zhao}, \citenamefont {Yang},
  \citenamefont {Chen}, \citenamefont {Zhou},\ and\ \citenamefont
  {Qiao}}]{liuFullyControllablePancharatnamBerry2016}%
  \BibitemOpen
  \bibfield  {author} {\bibinfo {author} {\bibfnamefont {Chuanbao}\
  \bibnamefont {Liu}}, \bibinfo {author} {\bibfnamefont {Yang}\ \bibnamefont
  {Bai}}, \bibinfo {author} {\bibfnamefont {Qian}\ \bibnamefont {Zhao}},
  \bibinfo {author} {\bibfnamefont {Yihao}\ \bibnamefont {Yang}}, \bibinfo
  {author} {\bibfnamefont {Hongsheng}\ \bibnamefont {Chen}}, \bibinfo {author}
  {\bibfnamefont {Ji}~\bibnamefont {Zhou}}, \ and\ \bibinfo {author}
  {\bibfnamefont {Lijie}\ \bibnamefont {Qiao}},\ }\bibfield  {title} {\enquote
  {\bibinfo {title} {Fully {{Controllable Pancharatnam}}-{{Berry Metasurface
  Array}} with {{High Conversion Efficiency}} and {{Broad Bandwidth}}},}\
  }\href {\doibase 10.1038/srep34819} {\bibfield  {journal} {\bibinfo
  {journal} {Scientific Reports}\ }\textbf {\bibinfo {volume} {6}} (\bibinfo
  {year} {2016}),\ 10.1038/srep34819}\BibitemShut {NoStop}%
\bibitem [{\citenamefont {Pan}\ \emph {et~al.}(2018)\citenamefont {Pan},
  \citenamefont {Cai}, \citenamefont {Tang}, \citenamefont {Zhou},\ and\
  \citenamefont {Dong}}]{panTrifunctionalMetasurfacesConcept2018}%
  \BibitemOpen
  \bibfield  {author} {\bibinfo {author} {\bibfnamefont {Weikang}\ \bibnamefont
  {Pan}}, \bibinfo {author} {\bibfnamefont {Tong}\ \bibnamefont {Cai}},
  \bibinfo {author} {\bibfnamefont {Shiwei}\ \bibnamefont {Tang}}, \bibinfo
  {author} {\bibfnamefont {Lei}\ \bibnamefont {Zhou}}, \ and\ \bibinfo {author}
  {\bibfnamefont {Jianfeng}\ \bibnamefont {Dong}},\ }\bibfield  {title}
  {\enquote {\bibinfo {title} {Trifunctional metasurfaces: Concept and
  characterizations},}\ }\href {\doibase 10.1364/OE.26.017447} {\bibfield
  {journal} {\bibinfo  {journal} {Optics Express}\ }\textbf {\bibinfo {volume}
  {26}},\ \bibinfo {pages} {17447} (\bibinfo {year} {2018})}\BibitemShut
  {NoStop}%
\bibitem [{\citenamefont {Jiang}\ \emph {et~al.}(2018)\citenamefont {Jiang},
  \citenamefont {Chen}, \citenamefont {Zhang},\ and\ \citenamefont
  {Chen}}]{jiangAchromaticElectromagneticMetasurface2018}%
  \BibitemOpen
  \bibfield  {author} {\bibinfo {author} {\bibfnamefont {Shan}\ \bibnamefont
  {Jiang}}, \bibinfo {author} {\bibfnamefont {Chang}\ \bibnamefont {Chen}},
  \bibinfo {author} {\bibfnamefont {Hualiang}\ \bibnamefont {Zhang}}, \ and\
  \bibinfo {author} {\bibfnamefont {Weidong}\ \bibnamefont {Chen}},\ }\bibfield
   {title} {\enquote {\bibinfo {title} {Achromatic electromagnetic metasurface
  for generating a vortex wave with orbital angular momentum ({{OAM}})},}\
  }\href {\doibase 10.1364/OE.26.006466} {\bibfield  {journal} {\bibinfo
  {journal} {Optics Express}\ }\textbf {\bibinfo {volume} {26}},\ \bibinfo
  {pages} {6466} (\bibinfo {year} {2018})}\BibitemShut {NoStop}%
\bibitem [{\citenamefont {Pozar}(2011)}]{pozarMicrowaveEngineering4th2011}%
  \BibitemOpen
  \bibfield  {author} {\bibinfo {author} {\bibfnamefont {David~M.}\
  \bibnamefont {Pozar}},\ }\href@noop {} {\emph {\bibinfo {title} {Microwave
  {{Engineering}} 4th {{Edition}}}}}\ (\bibinfo  {publisher} {{Wiley}},\
  \bibinfo {year} {2011})\BibitemShut {NoStop}%
\bibitem [{\citenamefont {Selvanayagam}\ and\ \citenamefont
  {Eleftheriades}(2016)}]{SelvanayagamDesignMeasurementTensor2016}%
  \BibitemOpen
  \bibfield  {author} {\bibinfo {author} {\bibfnamefont {M.}~\bibnamefont
  {Selvanayagam}}\ and\ \bibinfo {author} {\bibfnamefont {G.V.}\ \bibnamefont
  {Eleftheriades}},\ }\bibfield  {title} {\enquote {\bibinfo {title} {Design
  {{And Measurement}} of {{Tensor Impedance Transmitarrays For Chiral
  Polarization Control}}},}\ }\href {\doibase 10.1109/TMTT.2015.2505718}
  {\bibfield  {journal} {\bibinfo  {journal} {IEEE Transactions on Microwave
  Theory and Techniques}\ }\textbf {\bibinfo {volume} {64}},\ \bibinfo {pages}
  {414--428} (\bibinfo {year} {2016})}\BibitemShut {NoStop}%
\bibitem [{\citenamefont {Hannam}\ \emph {et~al.}(2014)\citenamefont {Hannam},
  \citenamefont {Powell}, \citenamefont {Shadrivov},\ and\ \citenamefont
  {Kivshar}}]{Hannam2014}%
  \BibitemOpen
  \bibfield  {author} {\bibinfo {author} {\bibfnamefont {Kirsty}\ \bibnamefont
  {Hannam}}, \bibinfo {author} {\bibfnamefont {David~A.}\ \bibnamefont
  {Powell}}, \bibinfo {author} {\bibfnamefont {Ilya~V.}\ \bibnamefont
  {Shadrivov}}, \ and\ \bibinfo {author} {\bibfnamefont {Yuri~S.}\ \bibnamefont
  {Kivshar}},\ }\bibfield  {title} {\enquote {\bibinfo {title} {Broadband
  chiral metamaterials with large optical activity},}\ }\href {\doibase
  10.1103/PhysRevB.89.125105} {\bibfield  {journal} {\bibinfo  {journal}
  {Physical Review B}\ }\textbf {\bibinfo {volume} {89}},\ \bibinfo {pages}
  {125105} (\bibinfo {year} {2014})}\BibitemShut {NoStop}%
\bibitem [{\citenamefont {Studio}(2018)}]{cst}%
  \BibitemOpen
  \bibfield  {author} {\bibinfo {author} {\bibfnamefont {CST~Microwave}\
  \bibnamefont {Studio}},\ }\href@noop {} {\enquote {\bibinfo {title}
  {Darmstadt - {Germany}},}\ } (\bibinfo {year} {2018})\BibitemShut {NoStop}%
\end{thebibliography}

\end{document}